\pdfoutput=1
\documentclass[12pt]{article}
\usepackage{graphicx}
\begin{document}

\title{GEOMETRIC VULNERABILITY OF DEMOCRATIC INSTITUTIONS AGAINST LOBBYING \\A sociophysics approach}

\author{SERGE GALAM\thanks{serge.galam@sciencespo.fr} \\ CEVIPOF - Centre for Political Research, \\ Sciences Po and CNRS,\\
 98 rue de l'Universit\'e Paris, 75007, France}
\date{August 10, 2016 \\ $\ $ \\(Mathematical Models and Methods in Applied Sciences, 2017)}

\maketitle

\begin{abstract}

An alternative voting scheme is proposed to fill the democratic gap between a president elected democratically via universal suffrage (deterministic outcome, the actual majority decides), and a president elected by one person randomly selected from the population (probabilistic outcome depending on respective supports). Indeed, moving from one voting agent to a group of $r$ randomly selected voting agents, reduces the probabilistic character of the outcome. Accordingly, building $r$ such groups, each one electing its president (elementary bricks), to constitute a group of the groups with the $r$ local presidents electing a higher-level president, does reduce further the outcome probabilistic aspect. The process is then repeated $n$ times to reach a bottom-up pyramidal structure with $n$ levels, $r^{n-1}$ elementary bricks at the bottom and a president at the top. Agents at the bottom are randomly selected but higher-level presidents are all designated according to the respective local majorities within the groups which elect them. At the top of the hierarchy the president is still elected with a probability but the distance from a deterministic outcome reduces quickly with increasing $n$. At a critical value $n_{c,r}$ the outcome turns deterministic recovering the same result a universal suffrage would yield. This alternative hierarchical scheme introduces several social advantages like the distribution of local power to the competing minority, which thus makes the structure more resilient, yet preserving the presidency allocation to the actual majority. It also produces an area around fifty percent for which the president is elected with an almost equiprobability slightly biased in favor of the actual majority. However, our results reveal the existence of a severe geometric vulnerability to lobbying. It is shown that a tiny lobbying group is able to kill the democratic balance by seizing the presidency democratically. It is sufficient to complete a correlated distribution of a few agents at the hierarchy bottom. Moreover, at the present stage, identifying an actual killing distribution is not feasible, which sheds a disturbing light on the devastating effect geometric lobbying can have on democratic hierarchical institutions.

\end{abstract}


\section{Introduction}	

\subsection{About sociophysics}

Sociophysics aims at a description of social and political behaviors using concepts and techniques from statistical physics. It should be stressed that it is neither a metaphoric description nor a biunique mapping of models from physics to social sciences. More precisely it is the physicist's approach which has to be implemented in tackling a social objet having in mind the way physicists dealt with the puzzles of inanimate matter in bulk to reach the powerful today accomplishments of statistical physics,\cite{book}. 

The instrumental key to implement a successful sociophysics modeling is to build specific models using tools and concepts from the physics of disorder. To transpose a model allocating social meaning to the corresponding physical variables may be a first step but should not be the last one.  At some point the social model must depart from its physical counterpart,\cite{review}.  That is the methodology I have been advocating and pushing forwards since the late seventies,\cite{book}. 

Is is  instructive to remind that in earlier eighties the physicist community was rather hostile to the new emerging field of sociophysics,\cite{testi}. I was rather lonely trying to convince my physicist colleagues focusing on the contradictory duality of their  situation with on the one hand, a totally closed horizon in terms of professional opportunities, and on the other hand, the incredibility powerful new tools they had mastered,\cite{revo,imper,frust}. In parallel I started to formalize my first contributions to the field publishing a series of innovative and founding papers,\cite{strike,entro,psy}.

Thirty years latter sociophysics, and more generally the formal modeling of social behavior, has became an active field of research among physicists,\cite{santo,bikas,bolek,cheon,weron-t,celia,koree,gerard,nuno,bagnoli,carbone,weron,zanette,iglesias,mobilla,fasano,andre,marcel}. With the development of multi-agent technics computer scientists are now joining the physicists quite naturally along social simulation,\cite{multi-a,marco}. More unexpected has been the new trend among mathematicians who started to get interested in sociophysics, uncovering new families of problems appropriate for mathematical investigation, e.g., this special issue,\cite{bello,lanchier}. Beside a few exceptions, social scientists are still very scarce to get involved in sociophysics,\cite{mosco}. However, time has come to confront sociophysical models to social sciences. Along this necessity, in 2013 I have joined the CEVIPOF, Centre for Political Research at Sciences Po, a leading social science institution in Paris. 

\subsection{Randomness and threshold dynamics}

Most sociophysical opinion models are found to obey threshold dynamics with tipping points and attractors. The update of agents is implemented using random selection and local rules. Depending on the model parameters and the diversity of agents, tipping points are often not symmetrical with respect to the competing choices. They can be found located at values as low as 15\% and as high as 85\% in case of two competing choices,\cite{mino}.  However, although the democratic balance may be heavily broken along one state, this very state still needs to get an initial support above the actual tipping point to invade the whole population. Randomness in the selection of agents is a key ingredient of these threshold dynamics. 

However in this paper I show that  threshold-like geometrical structures display the possibility for a tiny number of agents to democratically seize the top leadership with certaintly, despite the existence of  a tipping point to win. To bypass the threshold requirement demands to achieve a correlated geometrical nesting among these few agents. 

To illustrate this process I propose an alternative voting scheme to fill the democratic gap between a president elected democratically via universal suffrage and a president elected by one person randomly selected from the population. The outcome is deterministic in the first case with the actual majority within the whole population deciding on the president. Knowing the actual supports for the respective competing candidates allows to predict with certainty the voting outcome. It is only because these supports are not know precisely, even if polls can indicate good estimates at a given time, that the vote itself must take place in order to find out their respective values and thus the corresponding winner. On the contrary, the second case yields by definition a probabilistic outcome. Even knowing before end the different supports would not allow to foresee the actual outcome. Only the actual measure reveals the winner.

More precisely, for two competing candidates A and B with respective supports $p_0$ and $(1-p_0)$, universal suffrage leads with certainty to president A when $p_0>\frac{1}{2}$ and to president B otherwise ($p_0<\frac{1}{2}$). However a one person voting randomly selected from the entire population yields  president A with probability $p_0$ and president B with probability  $(1-p_0)$. Therefore, to move from one voting agent to a group of $r$ randomly selected voting agents, reduces the probabilistic character of the outcome. The new outcome probability $p_{1}$ arises from all configurations of $r$ agents with a majority of A agents satisfying $p_{1}>p_0$ when $p_0>\frac{1}{2}$ and $p_{1}<p_0$ for $p_0<\frac{1}{2}$.

This result hints at gathering together $r$ such groups, denoted elementary bricks, to have their respective presidents to elect a higher level president with a probability $p_{2}>p_{1}>p_0$ when $p_0>\frac{1}{2}$ and $p_{2}<p_{1}<p_0$ for $p_0<\frac{1}{2}$, thus reducing once more the outcome probabilistic aspect. Repeating the process $n$ times ends up in the building of a bottom-up pyramidal structure with $n$ levels, $r^{n-1}$ elementary bricks at the bottom and a president at the top.

Although agents are randomly selected at the bottom, higher-level presidents are all designated with certainty according to the respective local majorities within the groups which elect them. However, the hierarchy top president outcome is still given by a probability $p_{r,n}$ to account for all possible configurations of the $r^n$ agents. Nevertheless, the distance from a deterministic outcome reduces quickly with increasing $n$. Indeed, at a critical value $n_{c,r}$ the outcome turns deterministic recovering the same result a universal suffrage would give with $p_{n_{c,r}}\approx 1$ for $p_0>\frac{1}{2}$ and $p_{n_{c,r}}\approx 0$ for $p_0<\frac{1}{2}$,\cite{three,simu,general}.

This alternative hierarchical scheme introduces several social advantages like the distribution of local power to the competing minority, which thus makes the structure more resilient, yet preserving the presidency allocation to the actual majority. It also produces an area around fifty percent for which the president is elected with an almost equiprobability slightly biased in favor of the actual majority, thus avoiding hung election controversies,\cite{fifty}.

However, our results reveal the existence of a severe geometric vulnerability of these democratic bottom-up structures to lobbying. It is found that a tiny lobbying group is able to kill the democratic balance by seizing the presidency democratically from the occupation of a few targeted bottom slots.  It is sufficient to complete a correlated distribution of a few agents at the hierarchy bottom to reach the last level with certainty. Moreover, at the present stage, identifying an actual killing distribution is not feasible, which sheds a disturbing light on the devastating effect geometric lobbying can have on democratic hierarchical institutions.

\subsection{Outline}

The rest of the paper is organized as follows. Next Section analyzes the $r$ agent group voting election outcome when moving gradually from one agent to the entire population. Third Section investigates the process of circumventing the probabilistic outcome of the one voting group by building hierarchies. Conditions to recover a deterministic outcome are outlined in Section four. The existence of rare antidemocratic bottom configurations is studied in the fifth Section. The outcome of a hierarchy with $n$ levels and bricks of size $r$ is compared to the one group voting which includes $r^n$ agents in Section six. Section seven shows how to design killing geometries from the identification of rare antidemocratic configurations. The impossibility to detect an on going geometrical lobbying is discussed in Section 8 while last Section contains some conclusion.

\section{Randomness versus probability for presidential one group voting elections }	

Consider a population of $N$ agents, each one supporting either one of two competing political orientations $A$ or $B$ with respective proportions $p_0$ and  $1-p_0$. Although those proportions are not known they can be estimated at a given time more or less precisely using polls. While a president can be elected directly by the whole population using general elections, we address the question of alternative democratic elective schemes. However using randomly selected subsets of the population instead of a direct election shifts automatically the deterministic character of the election, the actual majority wins, into a probabilistic outcome driven by the random selection of the involved voting agents.

 \subsection{The presidential total group voting election}	

A direct general election with all agents voting yields  a deterministic outcome. A $A$ president is elected when $p_0>\frac{1}{2}$ and  the president is $B$ for $p_0<\frac{1}{2}$. It embodies the basic axiom of democratic voting where the president is elected by the current majority. It corresponds to the so called universal suffrage presidential election.

\begin{figure}
\centering
\includegraphics[width=.80\textwidth]{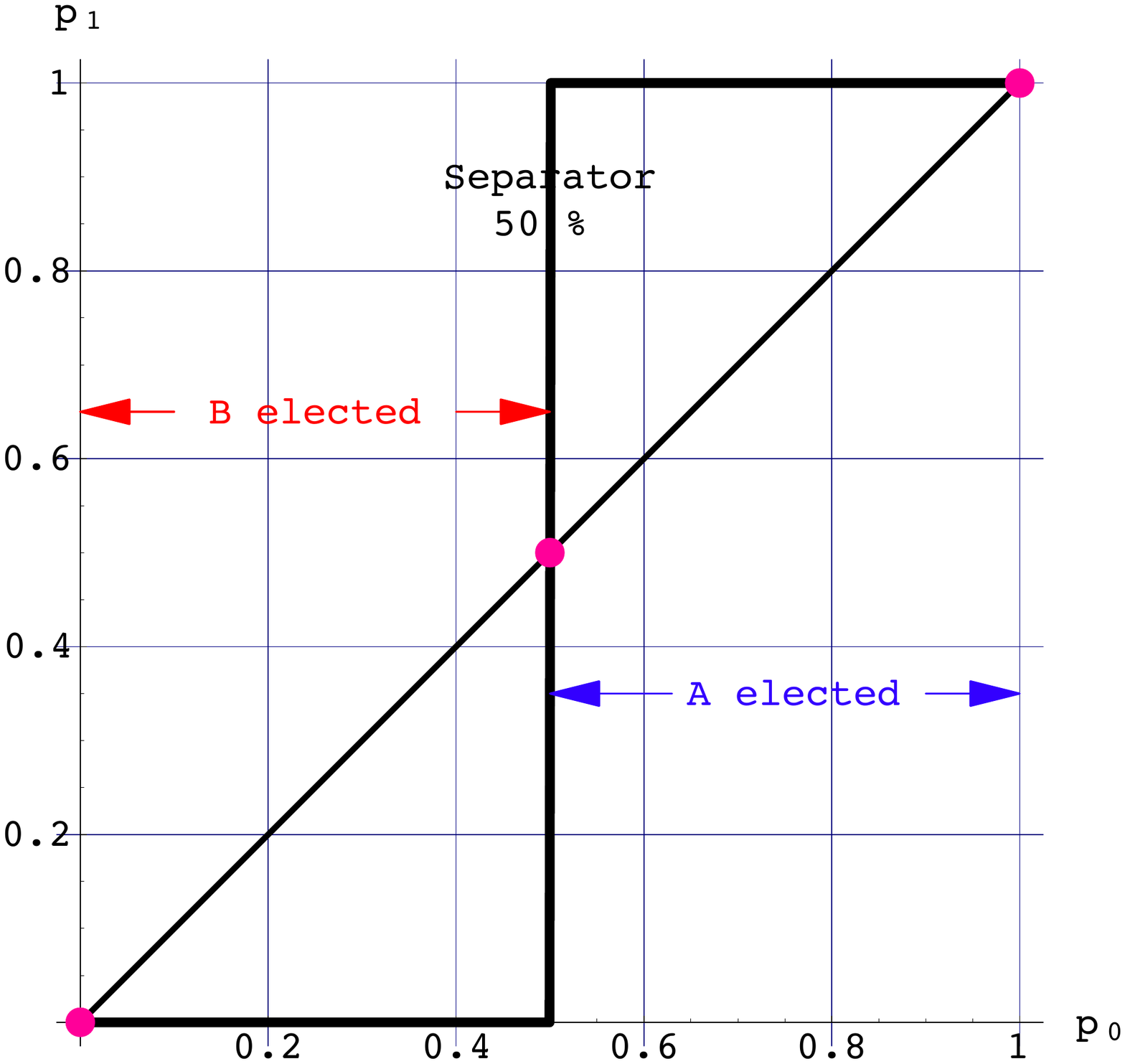}
\caption{The whole population voting step function (Eq. (\ref{p-all})). The result is deterministic, either yes or no depending on the density $p_0$ of A supporters. When $p_0>\frac{1}{2}$ an A is elected against a B for $p_0<\frac{1}{2}$.}
\label{p-all-a}
\end{figure}

\begin{figure}
\centering
\includegraphics[width=1.0\textwidth]{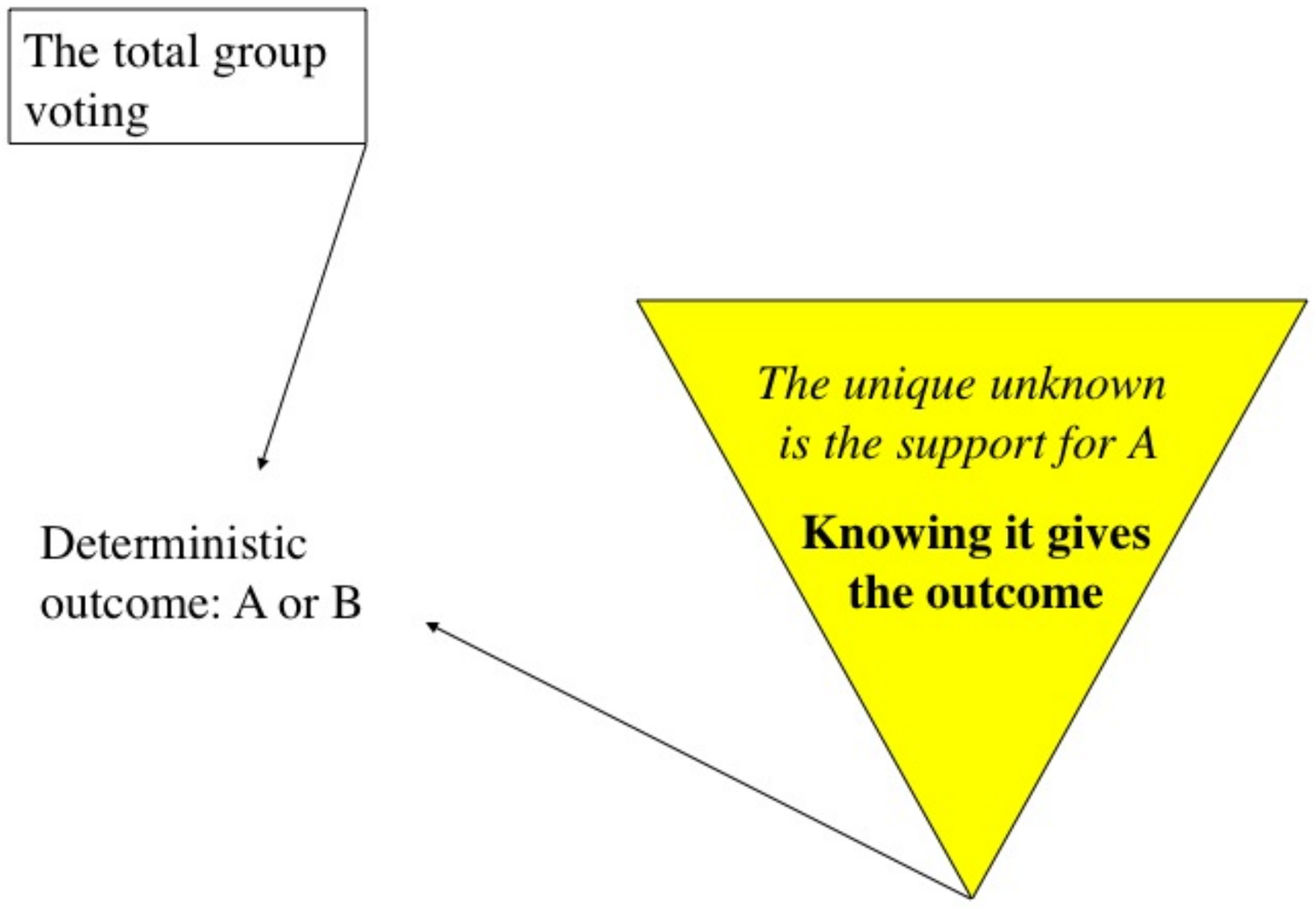}
\caption{The result for one vote in favor of $A$  from a group which includes the whole population with a density $p_0$ of A supporters and $(1-p_0)$ for $B$. The result is deterministic, either yes or no depending on $p_0$.}
\label{p-all-b}
\end{figure}

At this level we are considering a turnout of $100\%$. The outcome associated probability $p_1$ is a step function $P_{all}$ as
\begin{equation}
p_1\equiv P_{all}(p_0)=
\left\{
\begin{array}{cc}
 1 &   if\; p_0>1/2  \\
 0 &   if \; p_0<1/2.
\end{array} 
\right.
\label{p-all} 
\end{equation}
as exhibited in Figure (\ref{p-all-a}). It is total group voting where the group incorporates the whole population as explained in Figure (\ref{p-all-b}). 

Here we have neither randomness nor probability. The unique and essential unknown being the actual value of $p_0$.

 \subsection{The presidential one person group voting election}

\begin{figure}
\centering
\includegraphics[width=.80\textwidth]{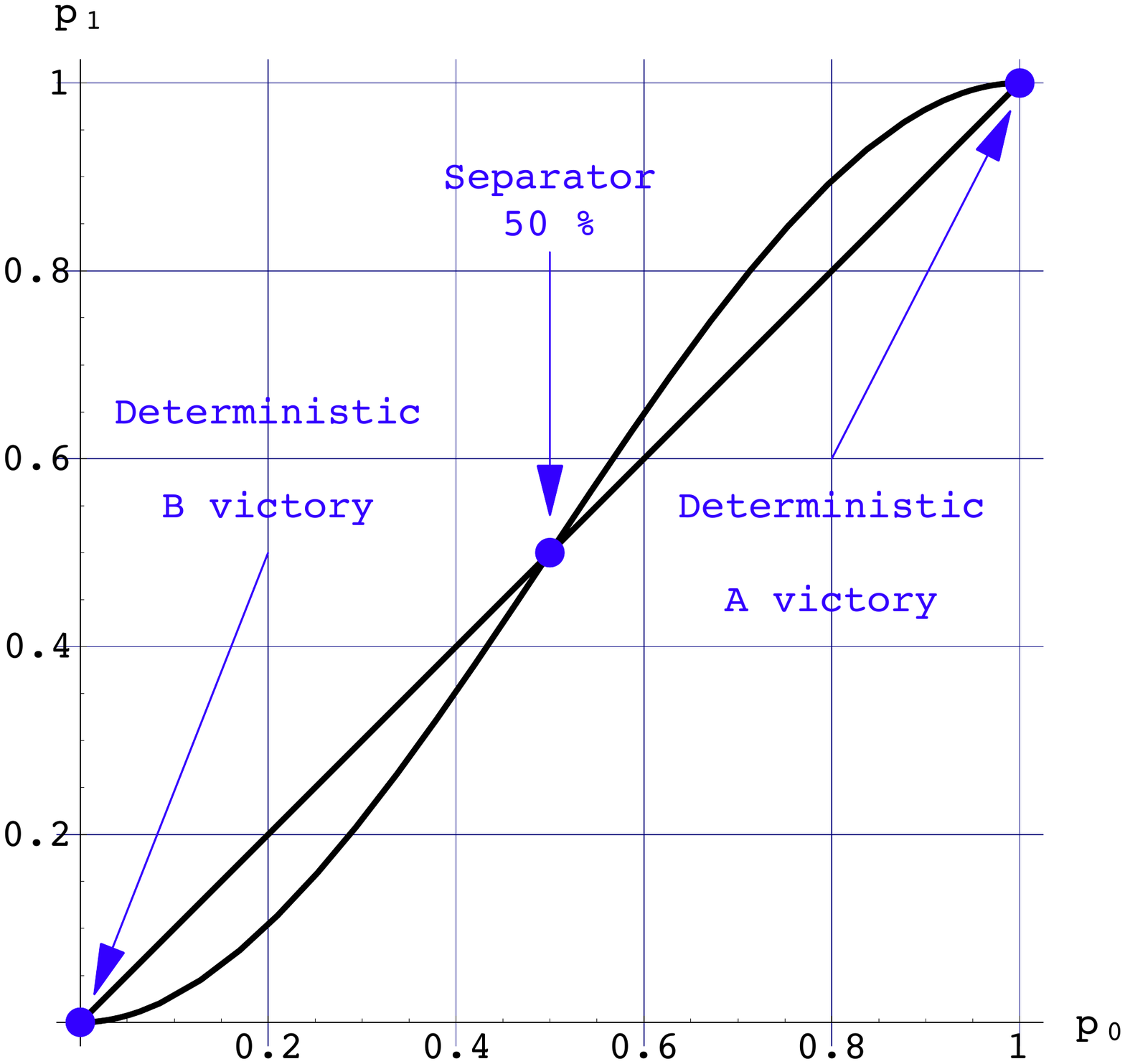}
\caption{The result for one vote in favor of $A$  from a one agent randomly selected from the population (straight diagonal line) and from a group of three agents randomly selected (curved line) from a population with a density $p_0$ of A supporters and $(1-p_0)$ for $B$. The result is now probabilistic in both case. For the one agent group $p_1=p_0$ while for the 3 agent group  $p_1<p_0$ when $p_0<\frac{1}{2}$ while  $p_1>p_0$ for $p_0>\frac{1}{2}$.}
\label{p3-a}
\end{figure}

At the opposite side of a total group voting one could consider a minimum group voting where one agent is selected randomly and votes for the president according to its own orientation either A or B. The outcome associated probability $p_1$ is a straight line $P_{1}$ as seen in Figure (\ref{p3-a}) with
\begin{equation}
p_1\equiv P_{1}(p_0)=p_0
\label{p-1} 
\end{equation}
which is a probabilistic outcome.
 
We now have both randomness and probability, the actual value of $p_0$ being still unknown. However, in case $p_0$ is known, the outcome is still probabilistic contrary to the total group voting election.

\subsection{The presidential three person group voting election}

\begin{figure}
\centering
\includegraphics[width=1.2\textwidth]{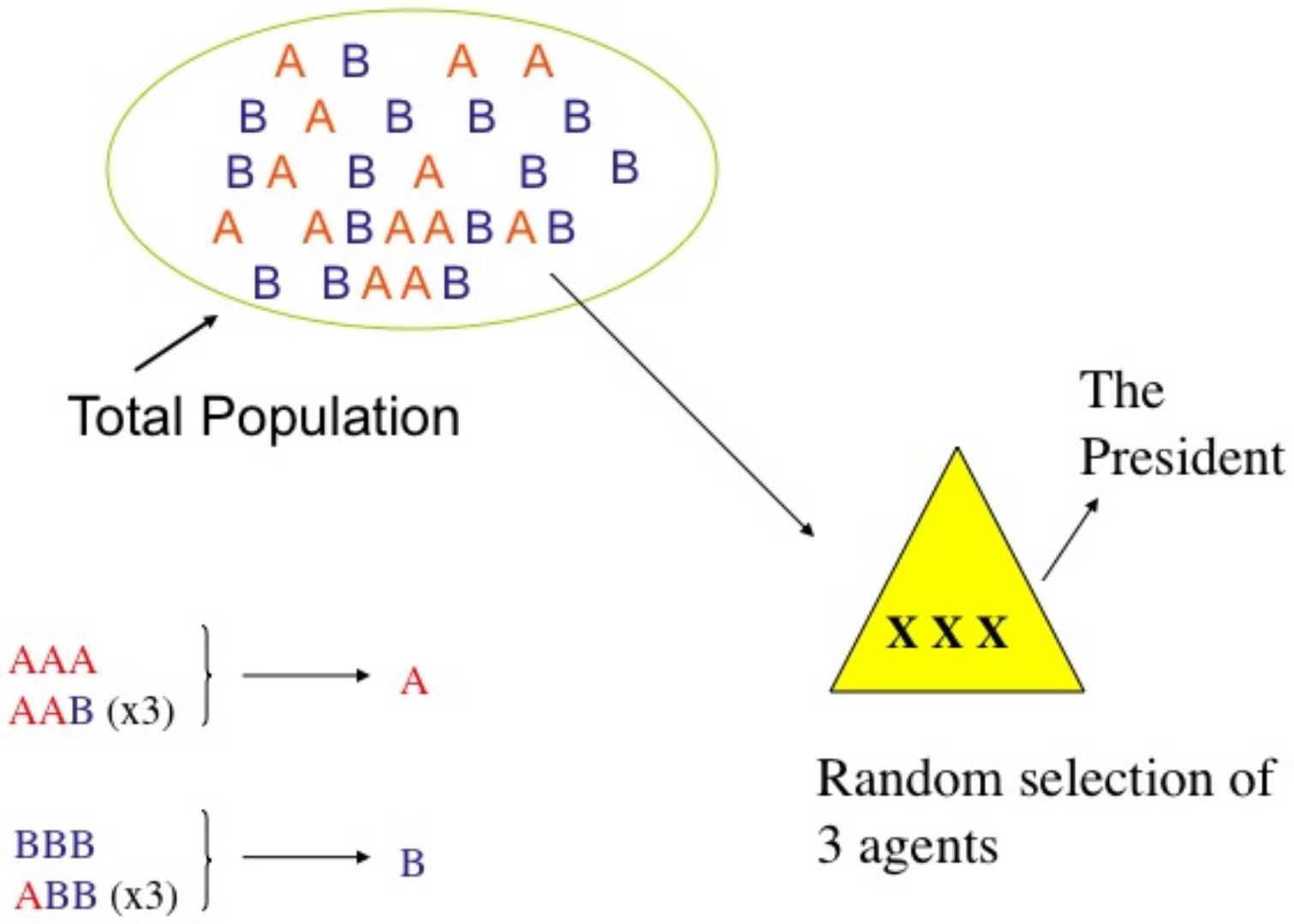}
\caption{Three agents are picked randomly from the population either A or B. The randomly selected group of $3$ agents elects the president.}
\label{random-3}
\end{figure}

\begin{figure}
\centering
\includegraphics[width=1.0\textwidth]{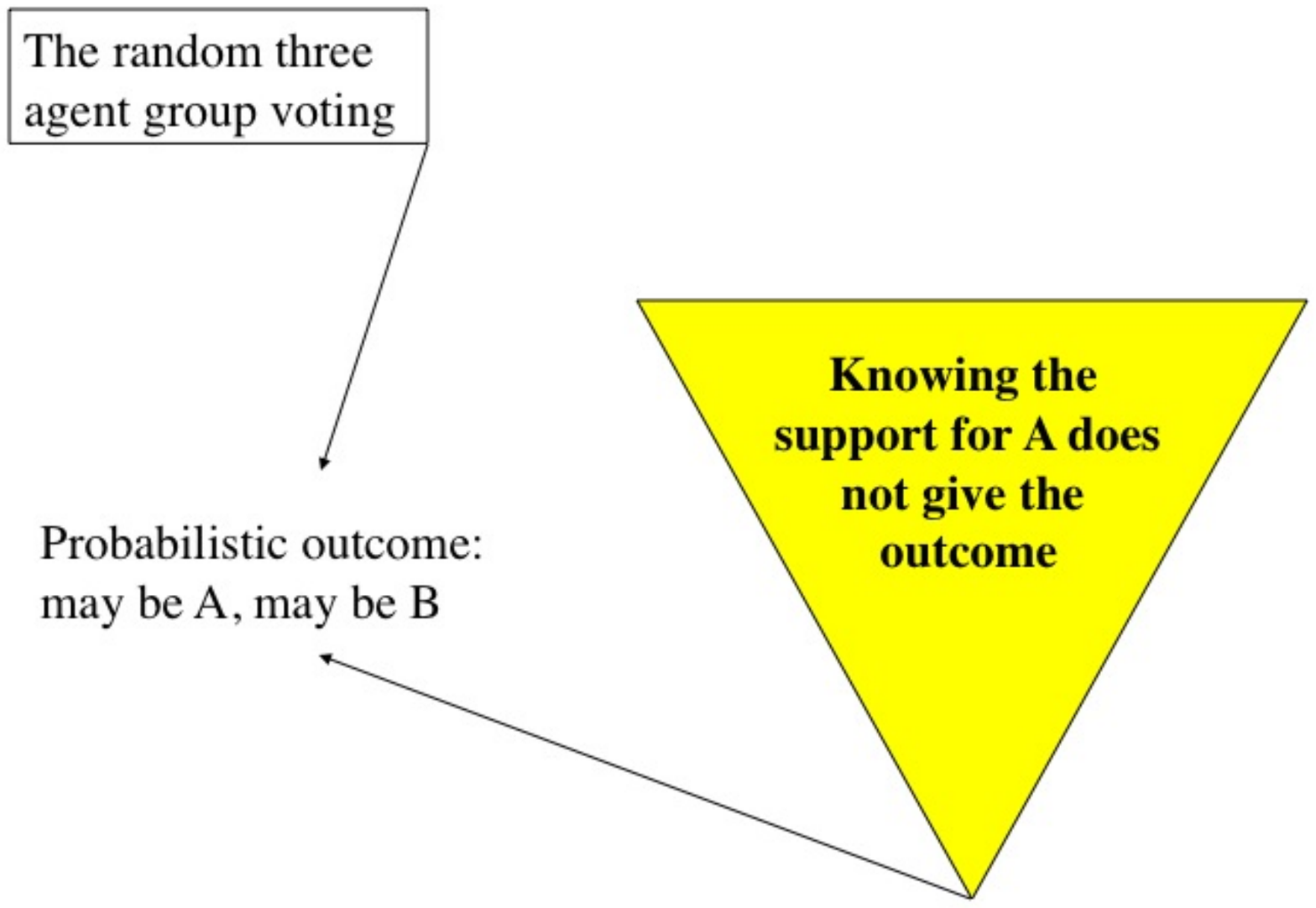}
\caption{The result for one vote in favor of $A$  from a group of three agents picked randomly from a population with a density $p_0$ of A supporters and $(1-p_0)$ for $B$. The result is probabilistic.}
\label{p3-b}
\end{figure}

Between above two extreme cases including either the full population or one single agent we can study the step by step construction of the total group. First step is to move from the one person group up to a three person group. Odd sizes are required to ensure the existence of a majority within the voting group. However, even sizes can be included provided eventual ties are broken randomly with equal probability among A and B. When three agents are randomly chosen from the population to elect the president (Fig.  (\ref{random-3})) $p_1$ is given  by a degree 3 polynomial function. A local $A$ majority  is given by either $\{3 A , 0 B\}$ or $\{2 A, 1 B\}$ yielding the probabilistic function
\begin{equation}
p_1\equiv P_3(p_0)=p_0^3+3 p_0^2 (1-p_0),
\label{p3} 
\end{equation}
shown in Fig. (\ref{p3-a}). While Eq. (\ref{p-all}) produces always a deterministic voting outcome in favor of the current majority, using Eq. (\ref{p3}) turns the process probabilistic for the whole range $0<p_0<1$ (Fig. (\ref{p3-b})). At  $p_0=0$  and $p_0=1$ the outcome is trivially deterministic. 

While $p_0=0.100$ yields the probability of $p_1=0.028$ to win the presidency, which is a rather improbable event, $p_0=0.300$ gives $p_1=0.216$, which is a likely event. However, a majority of $70\%$ has $78\%$ to win the election.  A short majority at $55\%$, gives only $58\%$ to get the presidency while an advantage of $5\%$ is huge in democratic societies.

Closer to $p_0=0.500$ the voting outcome becomes equivalent  to the flipping of a slightly imperfect coin. The coin bias being at the benefit of the actual majority. For instance $p_0=0.520$ wins only with $p_1=0.530$ at contrast to the $100\%$ from total group voting. The closer to $50\%$ the more equiprobable is the outcome with a high volatility in the actual selection of the three agents. 

In contrast, large differences like $90\%$ versus $10\%$ makes the majority to win at $97\%$, which is almost independent of the choices of the three voting agents in the sense that they are all likely to belong to the majority.

As for the one person group voting both randomness and probability are present no matter if the  the actual value of $p_0$ is known or not.

\subsection{The presidential $r$ person group voting election}

To preserve the current democratic balance the instrumental question is to ensure that the local majority within the selected group of $r$ agents is identical to the overall majority within the full population. We have seen that large differences in respective supports for A and B reduce the statistical fluctuations in the selection of the three agents. Configurations  which thwart the expected democratic balance with the current minority winning the election are very improbable. 

At the same time small differences in respective supports  is highly enhanced using three person group voting. In this case, statistical fluctuations associated to the random selection of three agents are amplified and to have a minority president becomes more likely.

From above size three group results we see that having a group of three agents instead of one reduces the distance to the deterministic outcome, which corresponds to the total group voting. If $p_0<\frac{1}{2}$ we get $p_1<p_0$, which means a value closer to zero, the associated democratic outcome. By symmetry, for $p_0>\frac{1}{2}$ we get $p_1>p_0$, which means a value closer to one, the corresponding democratic outcome.

It is thus interesting to study more quantitatively the connection between fluctuations, group sizes and democratic balance while going from the minimum size group with one single agent up to the maximum size group which includes the whole population. To evaluate the effect of the size voting group on the amplitude of the voting fluctuations given by the distance of $p_1$ to $100\%$, we consider the general $r$ person group voting case with the $r$ agents being still randomly selected from the population. 

A local majority is achieved with  configurations $\{r A, 0 B\}$, $\{(r-1) A, 1 B\}$, $\{(r-2) A, 2 B\}$... , $\{(\frac{r+1}{2}) A, \frac{r-1}{2}) B\}$. Accordingly, the corresponding probability $p_1=P_r(p_0)$ to have an $A$ elected by a group of $r$ agents randomly selected from the population writes
\begin{equation}
p_1= P_r(p_0)\equiv \sum_{m=\frac{r+1}{2}}^{r}  {r \choose m} p_0^m  (1-p_0)^{r-m},
\label{pr} 
\end{equation}
where $ {r \choose m}\equiv \frac{r !}{m ! (r-m) !}$ is a binomial coefficient. 

\begin{figure}
\centering
\includegraphics[width=.80\textwidth]{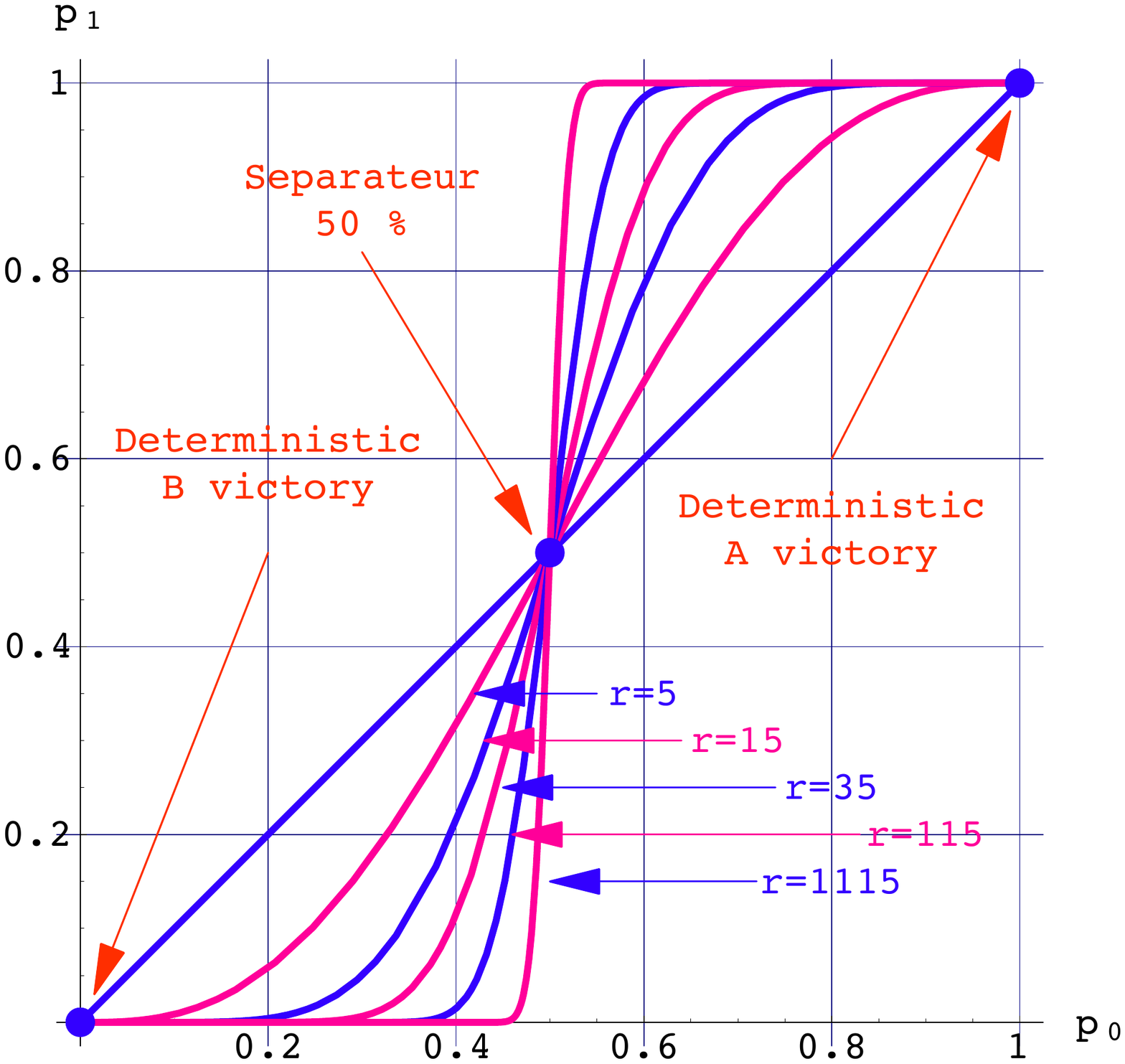}
\caption{Five different functions $P_r(p_0)$ are shown for $r=5, 15, 35, 115, 1115$. While the outcomes stay probabilistic for some range of $p_0$ they turn deterministic elsewhere. Increasing the value of $r$ reduces the range of value of $p_0$ which yields a probabilistic outcome.}
\label{pr-odd-a}
\end{figure}

The function $P_r(p_0)$ allows to estimate quantitatively the discrepancy between the outcome of an election using  $P_r$ versus $P_{all}$, which always yields a deterministic result. Voting outcomes associated to the values $r=5, 15, 35, 115, 1115$ are shown in Figure (\ref{pr-odd-a}) using Eq. (\ref{pr}). As seen from the Figure, increasing the value of $r$ reduces the range of value of $p_0$ which yields a probabilistic outcome with a large ranges of $p_0$ for which a quasi-determinisitc outcome is obtained.

For instance $p_0=0.300$ yields $p_1=0.163$ for a group of 5 agents against $p_1=0.216$ for the former case of 3 agents. This value falls down to $p_1=0.006$ for a group size $r=35$ and reaches $p_1=0.004$ already at a size $r=39$.  Rounding at two digits we can state that for a population of any size with two competing orientations with supports at respectively $30\%$ and $70\%$ a randomly selected group of only 39 agents yields the same outcome of electing a candidate from the majority of the full population, which could include millions of people.

The single random group voting can thus substitute itself to the total group voting provided the group size is tuned to recover a deterministic result. The problem being that  an increase of support at the benefit of the minority as to be damped by increasing the voting group size (Fig. 7).

Table (\ref{pr-1}) shows the various values of $p_1$ obtained for respectively $p_0=0.100, 0.200$, $0.300, 0.400$, $0.450, 0.470, 0.490$, $0.500, 0.510, 0.530$, $0.550,0.600$, $0.700, 0.800, 0.900$ using for each value $r=5, 15, 35, 115, 1115, all$. In the range $p_0<0.47$ or $p_0>0.53$, the $r=all$ result is always recovered. When $0.47<p_0<0.53$ the size $r=1115$ is not sufficient to overcome the fluctuations. However larger size group will do as for instance $p_0=0.470$ at $r=2000$ yielding $p_1=0.004$.

\begin{table}
\centering
\caption{ The single group voting outcome $p_1$ is displayed for respectively $p_0=0.100, 0.200 0.300, 0.400$, $0.450, 0.470, 0.490$, $0.500, 0.510, 0.530$, $0.550,0.600, 0.700, 0.800, 0.900$ using $r=5, 15, 35, 115, 1115, all$.}
\label{pr-1}   
\begin{tabular}{lllllll}
\hline\noalign{\smallskip}
$p_0 $  & $p_1$ & $p_1$ &  $p_1$ &  $p_1$ &  $p_1$ &  $p_1$   \\
$ $  & $r=3$ & $r=5$ &  $r=35$ &  $r=115$ &  $ r=1115$ &  $ r=all$ \\
\noalign{\smallskip}\hline\noalign{\smallskip}
0.100 & 0.028 & 0.009 & 0.000 & 0.000 & 0.000 & 0\\ \hline
0.200 & 0.104 & 0.058 & 0.000 & 0.000 & 0.000 & 0\\ \hline
0.300 & 0.216 & 0.163 & 0.006 & 0.000 & 0.000 & 0\\ \hline
0.400 & 0.352 & 0.317 & 0.114 & 0.015 & 0.000 & 0\\ \hline
0.450 & 0.425 & 0.407 & 0.275 & 0.141 & 0.000 & 0\\ \hline\hline\hline
0.470 & 0.455 & 0.444 & 0.360 & 0.259 & 0.022 & 0\\ \hline
0.490 & 0.485 & 0.481 & 0.453 & 0.415 & 0.252 & 0\\ \hline\hline
0.500 & 0.500 & 0.500 & 0.500 & 0.500 & 0.500 & 0.500\\ \hline\hline
0.510 & 0.515 & 0.519 & 0.547 & 0.585 & 0.745 & 1\\ \hline
0.530 & 0.545 & 0.556 & 0.640 & 0.741 & 0.978 & 1\\ \hline\hline\hline
0.550 & 0.575 & 0.593 & 0.743 & 0.859 & 1.000 & 1\\ \hline
0.600 & 0.648 & 0.683 & 0.886 & 0.985 & 1.000 & 1\\ \hline
0.700 & 0.784 & 0.837 & 0.994 & 1.000 & 1.000 & 1\\ \hline
0.800 & 0.896 & 0.942 & 1.000 & 1.000 & 1.000 & 1\\ \hline
0.900 & 0.972 & 0.991 & 1.000 & 1.000 & 1.000 & 1\\ \hline
\end{tabular}
\end{table}

\begin{figure}
\centering
\includegraphics[width=1.0\textwidth]{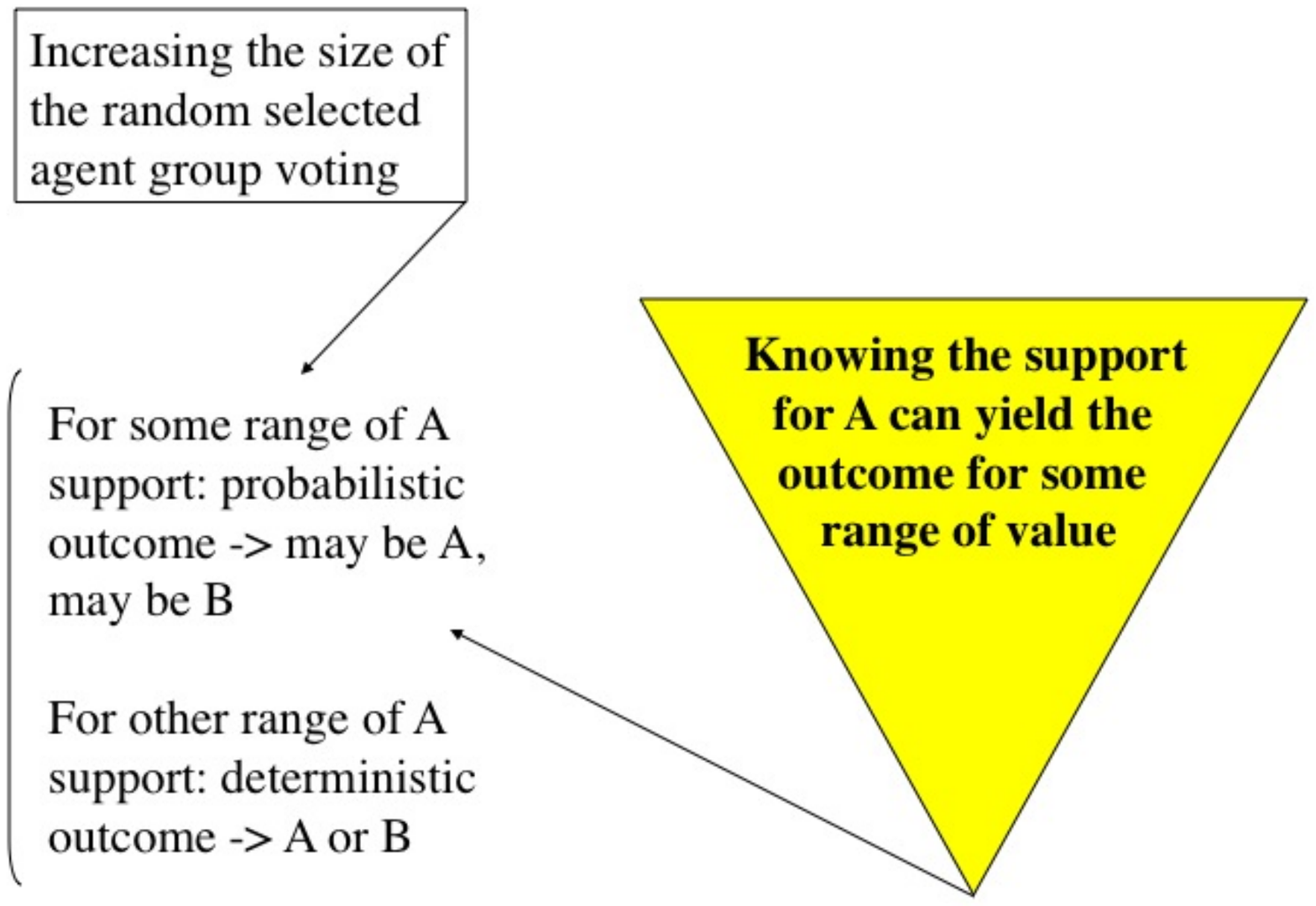}
\caption{Increasing the size of the random voting group from $r=3$ produces a range of values of $p_0$ where the outcome becomes deterministic as for the whole population voting group. Elsewhere the outcome is still probabilistic.}
\label{pr-odd-b}
\end{figure}

\subsection{The even size group voting election}	

Even sizes can also be included in the one group voting election. But to ensure a president is always elected preserving  he democratic majority rule,  one way is to break randomly eventual ties with equal probability at the benefit of A and B. Indeed with such a balanced tie break a group of even size $r=2l$ is identical to an odd size group with $r=2l-1$ as demonstrated below.

For any size $r$ both odd and even the voting function $P_r$  from Eq. (\ref{pr}) has to reformulated as
\begin{equation}
p_1= P_r(p_0)\equiv \sum_{m=N[ \frac{r}{2} ]}^{r}  {r \choose m} p_0^m  (1-p_0)^{r-m}
+k \delta\left \{\frac{r}{2}-N\left [\frac{r}{2}\right ] \right \}{r \choose \frac{r}{2}} p_0^\frac{r}{2}  (1-p_0)^\frac{r}{2},
\label{pr-odd-even} 
\end{equation}
where ${r \choose m}\equiv \frac{r!}{m!(r-m)!}$, $k$ is an integer with $0\leq k \leq 1$, $N[x]\equiv Integer\  part$ of  $x$ and $\delta\{x\}$ is the Kronecker function, i.e., $\delta\{x\}=1$ if $x=0$ and $\delta\{x\}=0$ if $x\neq 0$. The factor $k$ in the second part of Eq. (\ref{pr-odd-even}) allows to break the symmetry between both orientations  at an even group tie with a probability $k$ to contribute to A and $(1-k)$ for B.  

Any odd size $r$ yields $\{\frac{r}{2}-N[\frac{r}{2}]\}=\frac{1}{2}$ making the Kronecker function always equals to zero in Eq. (\ref{pr-odd-even}) thus canceling last term with Eq. (\ref{pr-odd-even}) becoming identical to Eq. (\ref{pr}).  For even $r$ size $\{\frac{r}{2}-N[\frac{r}{2}]\}=0$ with the last term contributing as $k {r \choose \frac{r}{2}} p_0^\frac{r}{2}  (1-p_0)^\frac{r}{2}$. Eq. (\ref{pr-odd-even}) can thus simply be written in two different expressions as
\begin{equation}
 P_r(p_0) \equiv  \sum_{m= \frac{r+1}{2} }^{r}  {r \choose m} p_0^m  (1-p_0)^{r-m} ,
\label{pr-odd} 
\end{equation}
for odd sizes $r$, and
\begin{equation}
 P_r(p_0)\equiv \sum_{m= \frac{r}{2} +1}^{r}  {r \choose m} p_0^m  (1-p_0)^{r-m}
+k {r \choose \frac{r}{2}} p_0^\frac{r}{2}  (1-p_0)^\frac{r}{2},
\label{pr-even} 
\end{equation}
when $r$ is even.

While $k\neq \frac{1}{2}$ produces very interesting results including quite naturally the effects of prejudices, collectives beliefs and cognitive bias in the process of decision making,\cite{mino,hetero} here no advantage is given to neither one of the two competing politics taking $k=\frac{1}{2}$. In such a case it is worth to emphasize that Eq. (\ref{pr-even}) with $r=2l$ is identical to Eq. (\ref{pr-odd}) with $r=2l-1$. In other terms 
\begin{equation}
P_{2l}(p_0)=P_{2l-1}(p_0) ,
\label{even=odd} 
\end{equation}
using Eq. (\ref{pr-odd-even}) with $k=\frac{1}{2}$. But as soon as  $k\neq \frac{1}{2}$ we have $P_{2l}(p_0) \neq P_{2l-1}(p_0)$.

To illustrate Eq. (\ref{even=odd}) a voting group of size 4 with $k=\frac{1}{2}$ has a the probabilistic voting function
\begin{equation}
p_1\equiv P_4(p_0)=p_0^4+4 p_0^3 (1-p_0)+3p_0^2 (1-p_0)^2,
\label{p4} 
\end{equation}
which reduces to $-2p^3+3p^2$, the probabilistic size 3 voting function given by Eq. (\ref{p3}). The same identity holds for $r=6$ and $r=5$ and so on. Therefore in the following we consider only odd $r$ values  without any loss of generality.

\section{Democratic hierarchies to circumvent the one group voting probabilistic character}	

Above analysis showed that on the one hand, selecting a voting group of $r$ randomly selected agents yield a probability $p_1<p_0$ to have an $A$ elected when $p_0<\frac{1}{2}$ while $p_1>p_0$ if $p_0>\frac{1}{2}$. On the other hand, increasing $r$ reduces the distance to the deterministic outcome of either zero or one obtained from the total group voting. 

\subsection{The elementary democratic probabilistic voting brick}

The distance of $\mid p_1-\frac{1}{2} \mid$ being always larger than  $\mid p_o-\frac{1}{2} \mid$ hints at considering a series of one $r$ voting group to produce a population of elected representative from which a new $r$ voting group could be formed to elect the president and thus reduces again the distance to the tipping point $\frac{1}{2}$.

To implement the scheme we consider the building of elementary presidential $r$ person voting groups which now elect respectively a first level representative still using a local majority rule. These groups will serve as elementary bricks denoted $r$-$brick_1$.

\begin{figure}
\centering
\includegraphics[width=0.8\textwidth]{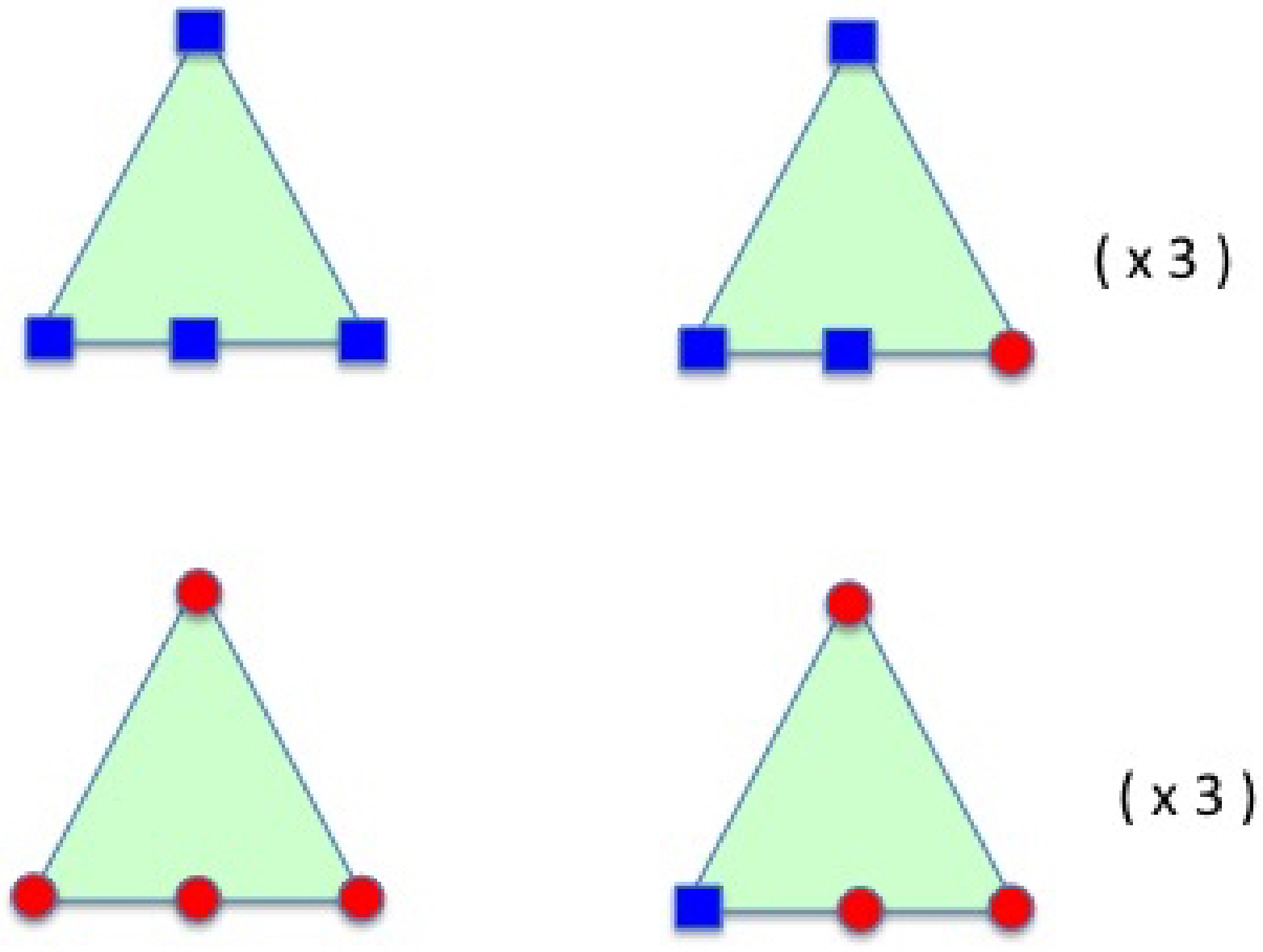}
\caption{The 8 elementary first level representative bricks ($3$-$brick_1$). A and B agents are represented respectively with red circles and blue squares. Upper brick yields a B elected while lower bricks have an A elected. Cases 2 against 1 (right column) exist in three different configurations with the minority agent at each of the three slots. }
\label{brick1}
\end{figure}

Each $r$ person voting group being formed by a random selection of the $r$ agent, a number $2^r$ different configurations are possible, half of them ($2^{r-1}$) yielding  majority A and the other half majority B. The case $r=3$ is shown in Figure (\ref{brick1}) with its associated $2^3=8$ different first level bricks, i.e., $3$-$brick_1$. Two bricks have respectively  3 A and 3 B, three have 2 A against 1 B and three 1 A against 2 B.

\subsection{The second level democratic probabilistic voting brick}

\begin{figure}
\centering
\includegraphics[width=0.8\textwidth]{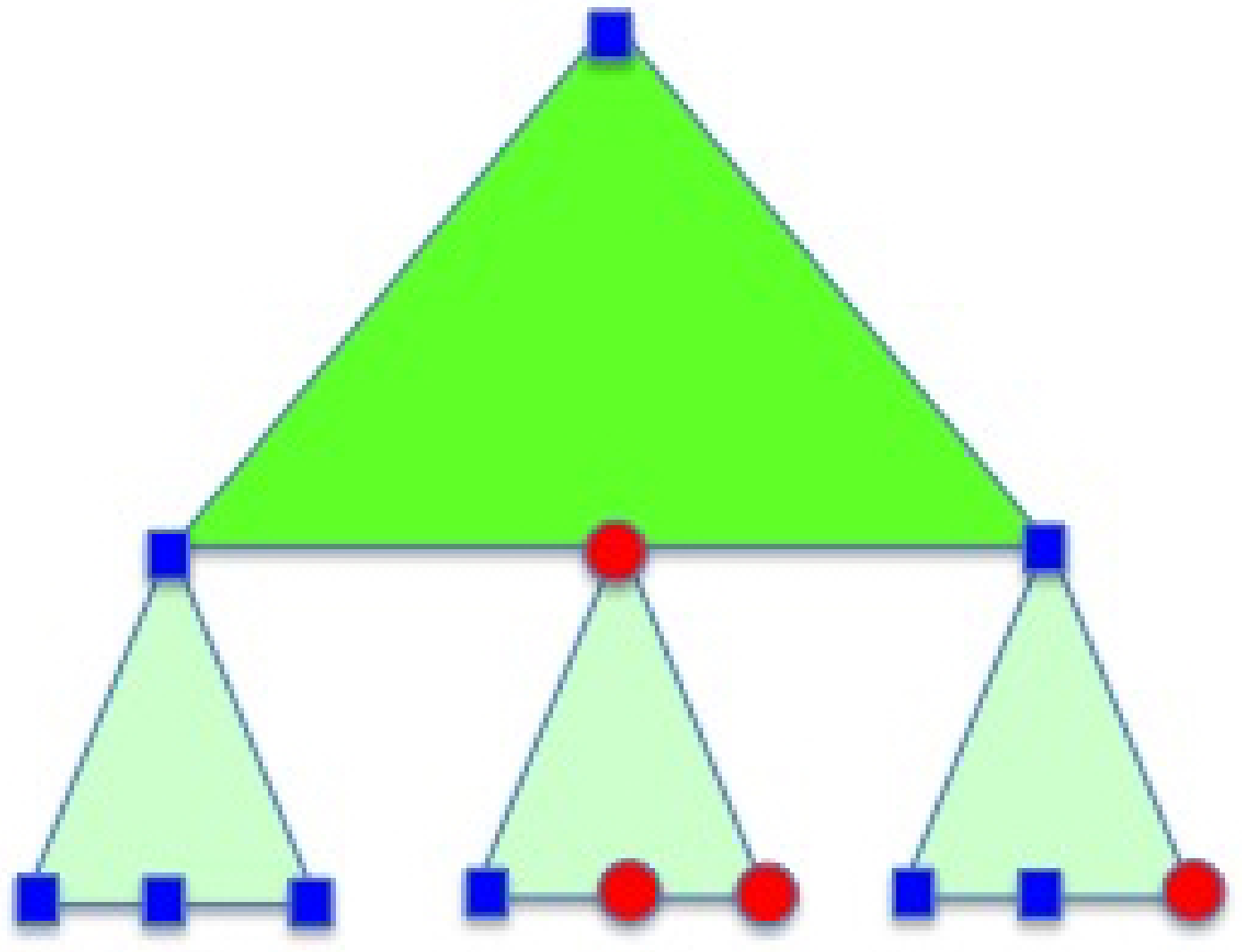}
\caption{Three first level representative bricks of size three ($3$-$brick_1$) are put together to form a $3$-$brick_2$. A and B agents are represented respectively with red circles and blue squares. Here two $3$-$brick_1$ (left and right) have B majority (respectively 3 B and \{1 A, 2 B\}) and one brick (middle) holding A majority \{ 2 A, 1 B\}. The  $3$-$brick_2$ yields president is B.}
\label{brick2}
\end{figure}

At this stage we can pick up randomly a series of $r$ elementary $r$-$brick_1$ to build a bigger second level elementary brick. The $r$ first level representatives form a new $r$ voting group which elects the president as exhibited in Figure (\ref{brick2}) for $r=3$. If the random selection of the $r$-$brick_1$ contains a majority of A first level representatives the president is A and it is B otherwise. 

Figure (\ref{brick2}) shows a case of  B majority with one peculiar series of $3$-$brick_1$. But for this very case there exists eight possible aggregations of $3$-$brick_1$. Four yield a majority of A first level elected representatives and four a B majority. Since the selection of the subset of three elementary $3$-$brick_1$ is random we can calculate the probability of having either an A or a B president. 

At the second level of the new two level brick the probability $p_2$ to get an A president is the sum of all the configurations with a majority ($\frac{r+1}{2}$) of A-$r$-$brick_1$. This sum is obtained simply by substituting $p_1$ to $p_0$ in Eq. (\ref{pr-odd}). We have
\begin{equation}
p_2= P_r(p_1)=P_r\{P_r(p_0)\},
\label{p2} 
\end{equation}
with inequalities  $\mid p_2-\frac{1}{2} \mid>\mid p_1-\frac{1}{2} \mid>\mid p_0-\frac{1}{2} \mid$.

\subsection{The $n$ level democratic probabilistic voting brick}

From Figure (\ref{brick2}) the resulting brick geometry appears to be pyramidal as expected by construction. But then,  since the distance to $\frac{1}{2}$ has been reduced twice in a row moving from the bottom level up to first and second level it is appealing to repeat the process a few more time to increase further the distance to $\frac{1}{2}$ thus reducing the probability to get a president belonging to the actual overall minority.

\begin{figure}
\centering
\includegraphics[width=0.8\textwidth]{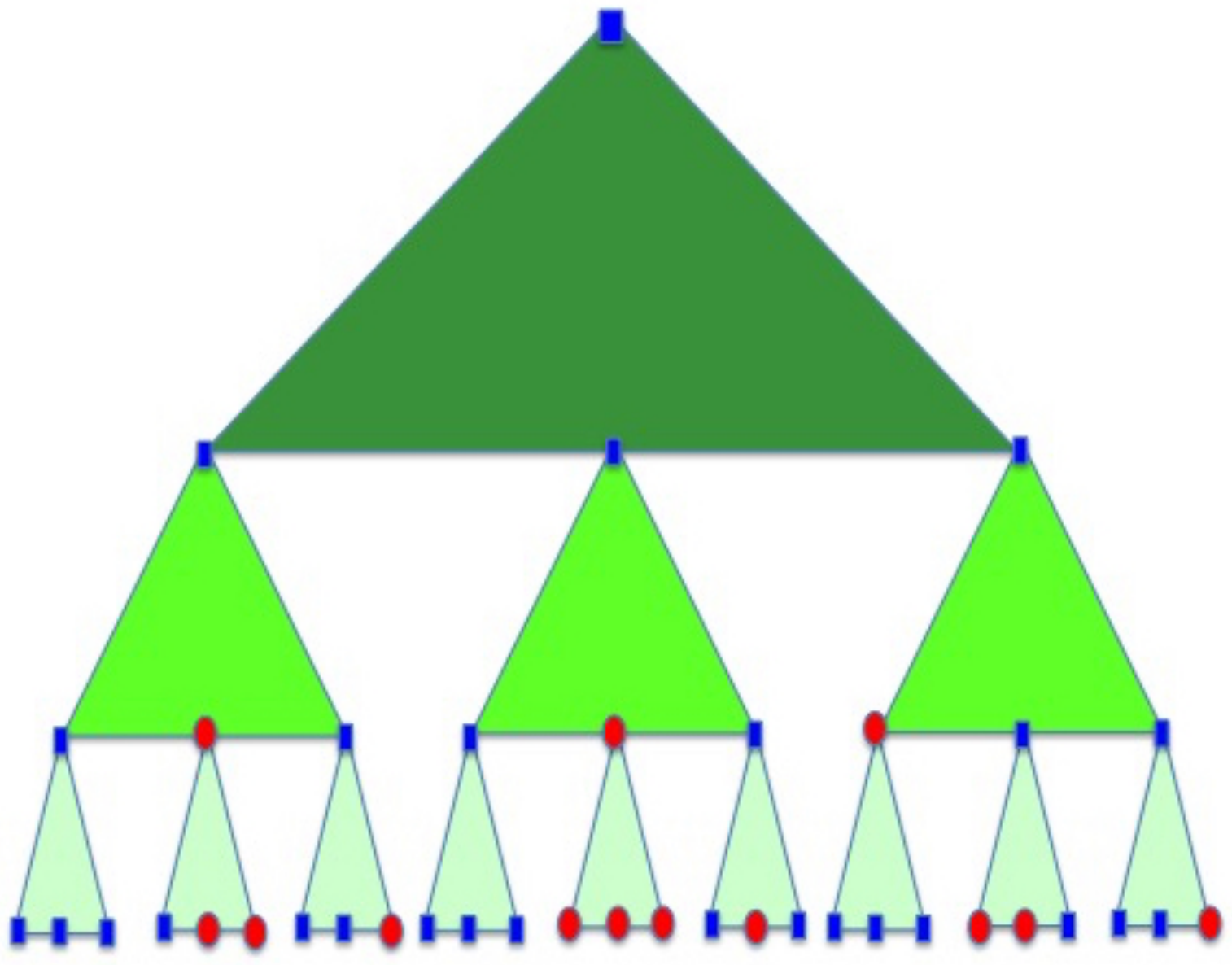}
\caption{Nine first level representative bricks of size three ($3$-$brick_1$) are put together to form a $3$-$brick_3$. A and B agents are represented respectively with red circles and blue squares. Six $3$-$brick_1$ with B majority and three with A majority are gathered to constitute a $3$-$brick_2$. The three $3$-$brick_2$ yield each B majority ending with a B president.}
\label{brick3}
\end{figure}

To build a $n$ level brick with voting groups of size $r$, i.e., a $r$-$brick_n$, $r^n$ groups of $r$ agents are required to constitute the first hierarchical level, i.,e., the bottom of the hierarchy. One case with $r=3$ and $n=3$ is shown in Figure (\ref{brick3}). It includes 9 groups of size three scoring to 27 agents at the bottom. The second level has 3 groups of 3 agents adding to 9. The third level has 1 group of  3 and at last level stands the elected president. The full hierarchy involves 40 agents of which 27 are randomly selected from the population. Others are chosen according to the respective votes of the various groups.

However, for most values $p_0$, the outcome is not deterministic with the probability $p_3$ being neither zero nor one as for instance with the series $p_0=0.400, p_1=0.352$, $p_2=0.284$ and $p_3=0.197$. In contrast an initial $p_0=0.200$ leads to $p_1=0.104$, $p_2=0.030$ and $p_3=0.000$.

At this stage the instrumental question is to determine if recovering a deterministic outcome (eventually equal to zero or one) is feasible within small values of $n$. To build a bottom-up democratic hierarchy with one hundred levels does not make sense. Given a $r$-$bricks_n$  hierarchy we want to determine the critical number of levels $n_{c,r}$ for which $p_{n_{c,r}}\approx 0$ for $p_0< \frac{1}{2}$ and $p_{n_{c,r}}\approx 1$ for $p_0> \frac{1}{2}$. 

\subsubsection{Fixed points and their stability}

Increasing the hierarchy level being identical to iterate Eq. (\ref{pr-odd}) we need first to study the associated dynamics by determining the corresponding fixed points solving $P_r(p*)=p*$. Three fixed points $p_B=0, p_c=\frac{1}{2}, p_A=1$ are found. At $p_B=0$ and $p_A=1$ any election is a trivial deterministic event with a hundred percent prediction of the outcome, respectively B and A. At $p_c=\frac{1}{2}$ the probability to have an A elected is equal to  the probability to have a B elected. It is worth to stress that for a group of size $r$, the voting function $P_r(p_0)$ is a polynomial of degree $r$ which has $r$ fixed points. However only $p_B, p_c, p_A$ satisfy the ``physical constraint" $0\leq p \leq 1$. 

To evaluate the respective stabilities of above three fixed points $p_B, p_c, p_A$ we expand the voting function $p_n=P_r(p_{n-1})$ around a fixed point $p*$ getting
\begin{equation}
p_n\approx p^{*}+(p_{n-1}-p^{*}) \lambda_r \ ,
\label{taylor-1/2}
\end{equation}
where $\lambda_r (p^*) \equiv \frac{dP_r(p_{n-1})}{dp_{n-1}}|_{p^{*}}$. 
Rewriting Eq. (\ref{taylor-1/2})  as
\begin{equation}
(p_n-p^{*})\approx (p_{n-1}-p^{*}) \lambda_r \ ,
\label{taylor}
\end{equation}
shows that $\lambda_r <1$ implies a stable fixed point while for $\lambda_r >1$ the fixed point is unstable. 

For the two extreme values $p_B=0$ and $p_A=1$ we have $\lambda_r(p_B)=\lambda_r(p_A)=0$ making both fixed points stable. 
For $p_c=\frac{1}{2}$ we get
\begin{equation}
\lambda_r(p_c)=\frac{1}{2^{r-1}}
\sum_{m=\frac{r+1}{2}}^{r} (2 m -r) {r \choose m}  \ ,
\end{equation}
which can be reduced to
 \begin{equation}
\lambda_r(p_c) =\frac{r}{2^{r-1}}  {r-1 \choose \frac{r-1}{2}} \ .
\label{lambda-1/2}
\end{equation}
Eq. (\ref{lambda-1/2}) yields $\lambda_r(p_c)>1$ for any size from
$r=3$ with $\lambda_3(p_c)=\frac{3}{2}$ to $\lambda_r(p_c) \rightarrow \sqrt \frac{2}{\pi}     \frac{r}{\sqrt (r-\frac{2}{3}}   \approx  \sqrt \frac{2r}{\pi} $ for 
$r \gg 1$ as seen in Table (\ref{lambda}).

\begin{table}
\centering
\caption{ Variation of $\lambda_r(p_c)$ as function the local group  size with  $r=3, 5, 7, 9,15, 101, 1001$. It is always larger than 1 making the fixed point $p_c=\frac{1}{2}$ unstable.}
\label{lambda}   
\begin{tabular}{llllllllll}
\hline\noalign{\smallskip}
 r & $ 3$ & $ 5$ &  $ 7$ &  $ 9$ &  $ 15$ & $ 101$ &  $ 1001$ \\
\noalign{\smallskip}\hline\noalign{\smallskip}
$\lambda_r(p_c)$ & 1.5 & 1.87 & 2.19 & 2.46 & 3.14 & 8.04 & 25.25 \\ \hline
\end{tabular}
\end{table}

Above results $\lambda_r(p_B)=\lambda_r(p_A)=0$ and $\lambda_r(p_c)>1$ show that the flow diagram of the dynamics driven by repeated voting exhibits two attractors located  at respectively $p_B=0$ and $p_A=1$, the separator being located at $p_c=\frac{1}{2}$. An initial support $p_0>\frac{1}{2}$ leads towards an election at $100\%$ provided enough voting levels are present. When $p_0<\frac{1}{2}$ it leads towards an election at $0\%$ for the A politics. 

\subsubsection{Critical number of hierarchical levels}

Having determined the flow diagram we can formally evaluate the number $n_{c,r}$ of hierarchical levels at which $p_{n_{c,r}}=0$ starting from $p_0< \frac{1}{2}$.
We iterate Eq. (\ref{taylor}) further from level $(n-1)$ to $(n-2)$ and so forth down to level 0 to get
\begin{equation}
(p_n-p_c)\approx (p_0-p_c) \lambda_r^n \ .
\label{taylor-n}
\end{equation}
At $n=n_{c,r}$ having by definition  $p_{n_{c,r}}=0$ turns Eq. (\ref{taylor-n}) to
\begin{equation}
 \lambda_r^{n_{c,r} }\approx \frac{p_c}{p_c-p_0}\ ,
\label{ta}
\end{equation}
from which, taking the Logarithm on both side yields
\begin{equation}
{n_{c,r}}\approx \frac{1}{\ln \lambda_r} \ln  \frac{1}{1-\frac{p_0}{p_c}}  \ .
\label{nc-}
\end{equation} 
When $p_0>p_c=\frac{1}{2}$ we get by symmetry
\begin{equation}
{n_{c,r}}\approx \frac{1}{\ln \lambda_r} \ln  \frac{1}{\frac{p_0}{p_c}-1}   \ .
\label{nc+}
\end{equation} 
where we used  $p_{n_{c,r}}=1$ instead of $p_{n_{c,r}}=0$. Eqs. (\ref{nc-}) and (\ref{nc+}) combine into the single Equation
\begin{equation}
{n_{c,r}}\approx  \frac{1}{\ln \lambda_r} \ln  \frac{1}{ \mid 1-\frac{p_0}{p_c} \mid } \ ,
\label{nc-+}
\end{equation}
which exhibits a logarithmic singularity at the unstable fixed point separator $p_c$. 

The singularity embodies the fact that when $p_0$ is very close to $p_c$ many  voting levels are necessary to increase the initial small difference in respective supports for A and B with volatility at its highest in the random selection of agents.

Table (\ref{nce}) compares the exact numerical estimates $n_{c,r}^e$ for $n_{c,r}$ obtained by successive iterations of Eq. (\ref{pr-odd}) to the values obtained from Eq. (\ref{nc-+}) in the case $p_0=0.49$  as a function of the series of voting size group $r=3, 5, 7, 9, 11, 13, 15$. 

The number of levels being by nature an integer Eq. (\ref{nc-+}) should be modified accordingly. Moreover, we notice that to match $n_{c,r}$ given  Eq. (\ref{nc-+}) to the exact estimates $n_{c,r}^e$ it is sufficient to take the integer part of Eq. (\ref{nc-+}) and to add $+2$ to the result besides for $r=3$. It leads  to the effective expression
\begin{equation}
\bar{n}_{c,r} \equiv N\left[ \frac{1}{\ln \lambda_r} \ln  \frac{1}{ \mid 1-\frac{p_0}{p_c} \mid }\right] +2 \ .
\label{nc-bis}
\end{equation}
This $+2$ addition has one $+1$ coming from taking the integer part of $n_{c,r}$ while the second $+1$ contribution is an ad hoc hand fit.

\begin{table}
\centering
\caption{Exact numerical estimates $n_{c,r}^e$ for $n_{c,r}$ using Eq. (\ref{pr-odd}) and the values obtained from Eq. (\ref{nc-+}) for the bottom value $p_0=0.49$ as a function of the series of voting size group $r=3, 5, 7, 9, 11, 13, 15$. Values $\bar{n}_{c,r}$ from Eq. (\ref{nc-bis}) are also given. }
\label{nce}   
\begin{tabular}{llllllll}
\hline\noalign{\smallskip}
$r$  & $ 3$ & $ 5$ &  $ 7$ &  $ 9$ &  $ 11$ &  $ 13$ &  $ 15$  \\
\noalign{\smallskip}\hline\noalign{\smallskip}
$n_{c,r}^e$ & 12 & 8 & 6 & 6 & 6 & 5 & 5 \\ \hline
$n_{c,r}    $ & 9.65  & 6.22 & 4.99 & 4.34 & 3.93 & 3.64 & 3.42 \\ \hline
$\bar{n}_{c,r}  $ & 11  & 8 & 6 & 6 & 5 & 5 & 5 \\ \hline
\end{tabular}
\end{table}

\begin{figure}
\centering
\includegraphics[width=.5\textwidth]{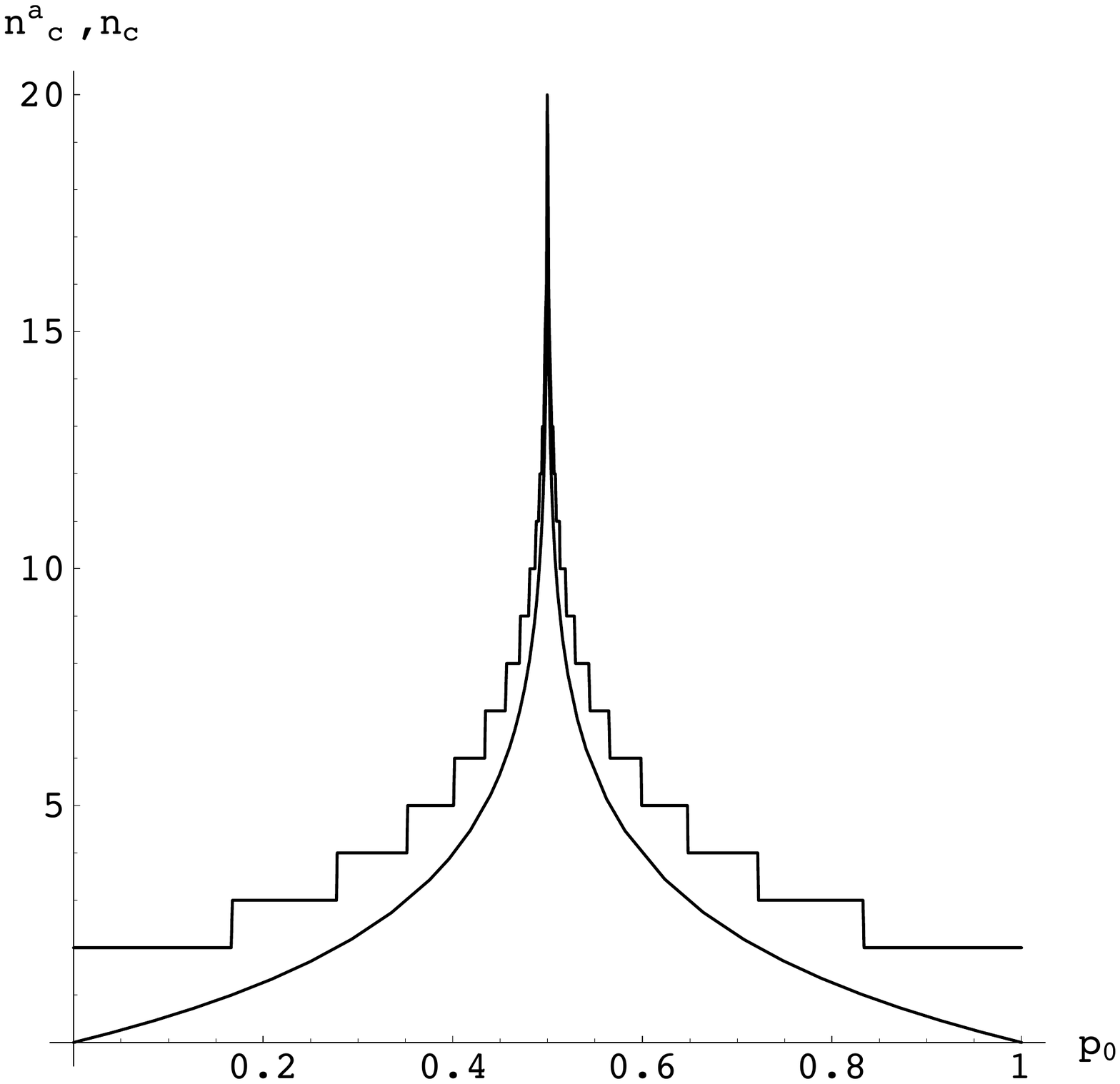}\hfill
\includegraphics[width=.5\textwidth]{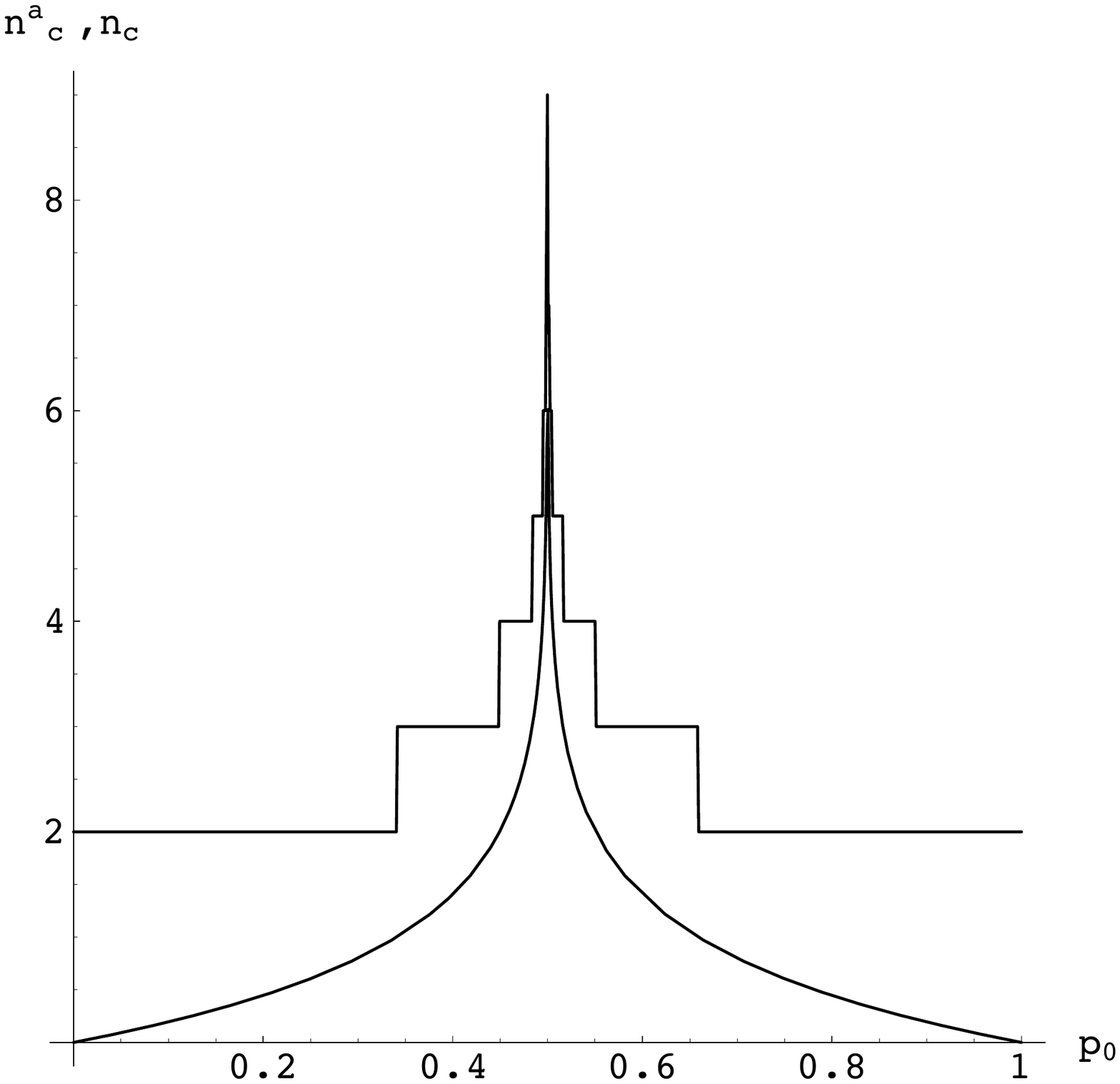}
\caption{The variation of $\bar{n}_{c,r}$ and $n_{c,r}$ for the two voting group sizes $r=3$ (left part) and $r=15$ (right part) as function of $0\leq p_0 \leq 1$. The stair functions correspond to $\bar{n}_{c,r}$. Both curves exhibit a logarithmic singularity at the unstable fixed point $p_c=\frac{1}{2}$ separator. Much less stairs appear for the larger size $r=15$.
}
\label{nc-3-15}
\end{figure}

Figure (\ref{nc-3-15}) illustrates the dependence of $\bar{n}_{c,r}$ and $n_{c,r}$ for the two voting group sizes $r=3$ and $r=15$ as function of $p_0$ with $0\leq p_0 \leq 1$.  The stair functions correspond to $\bar{n}_{c,r}$. Both curves shows a singularity at the separator $p_c=\frac{1}{2}$, which means that in the very vicinity of $50\%$ many voting levels are required to enlarge significantly the difference in respective supports while climbing up the hierarchy. Much less stairs appear for the larger size $r=15$ in agreement with the limit case of $r=all$, which yields $n_{c,r}=1$.

\section{Deterministic versus probabilistic outcome for a $n$-level hierarchy}

Given some support $p_0$ for A and $(1-p_0)$ for B a number $\bar{n}_{c,r}$ of bottom-up hierarchical levels was evaluated (Eq. (\ref{nc-bis})) to ensure the president ultimately elected at the top, i.e.,  the $\bar{n}_{c,r}$ level belongs to the overall majority in the population. However, $\bar{n}_{c,r}$ is a function of $p_0$ while in the social world political institutions do not change their structure at every election or decisional event. They have a fixed and constant number of hierarchical levels. On the other hand, the respective supports for competing A and B may vary quite subsequently from one election to another. This very fact discards the use of Eq. (\ref{nc-bis}) for any practical application to real situations. To make our analysis applicable the problem must be reformulated. 

The operational question is therefore to determine, given $n$ levels, what is the minimum overall support at the bottom if any, for either A or B to get the presidency with certainty. To answer the question we  invert Eq. (\ref{taylor-n}) to get
\begin{equation}
p_0=p_c+(p_n-p_c) \lambda_r^{-n} \ .
\label{inverse}
\end{equation}
Plugging $p_n=0$ yields the higher limit $p_{r,n,B}$  below which any vote produces a B victory at the top level. It defines a lower killing threshold below which A  gets a zero probability to win the presidency with
\begin{eqnarray}
p_{r,n,B} &=& p_c(1-\lambda_r^{-n})
\nonumber\\
&=& \frac{1}{2} (1-\lambda_r^{-n})  \ ,
\label{inverse1}
\end{eqnarray}
since $p_c=\frac{1}{2}$. In parallel plugging $p_n=1$ into Eq. (\ref{inverse}) gives a second threshold, the higher killing threshold $p_{r,n,A}$, above which the A gets  the top of the hierarchy with a total certainty with
\begin{eqnarray}
p_{r,n,A} &=& p_c(1-\lambda_r^{-n})+\lambda_r^{-n}
\nonumber\\
&=& \frac{1}{2} (1+\lambda_r^{-n})  \ .
\label{inverse2}
\end{eqnarray}

\begin{figure}
\centering
\includegraphics[width=.8\textwidth]{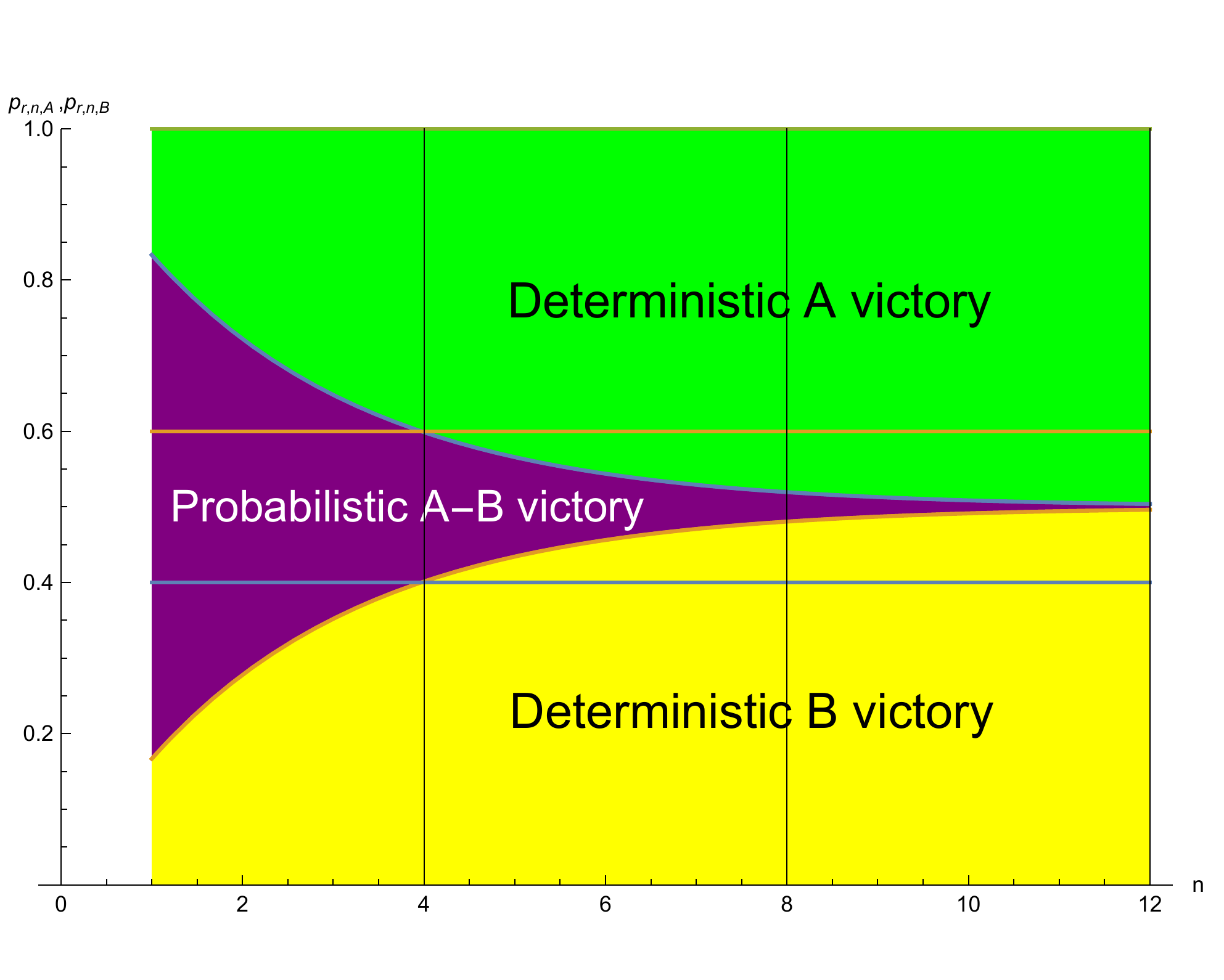}
\includegraphics[width=.8\textwidth]{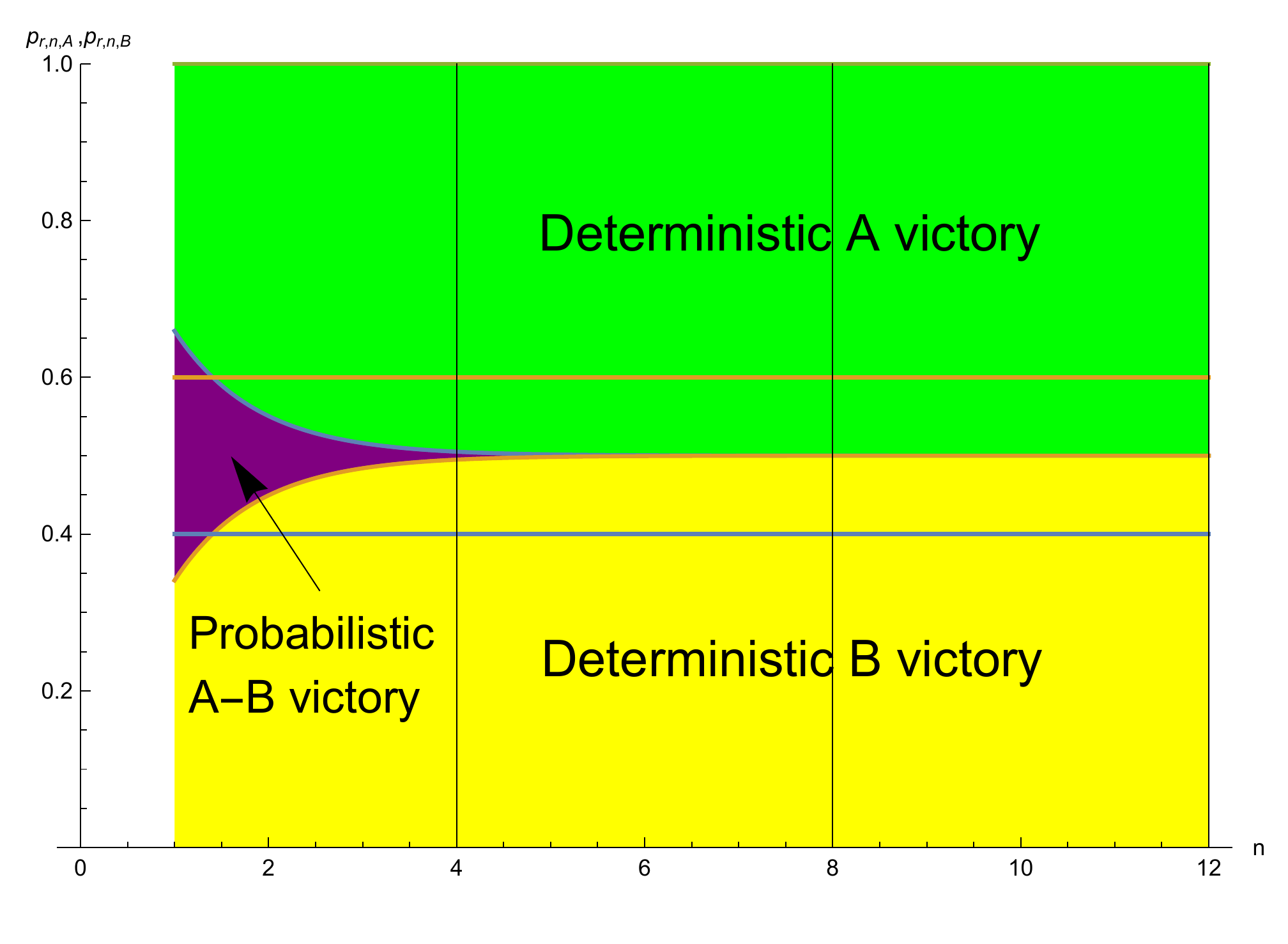}
\caption{Functions $p_{r,n,A}$ and $p_{r,n,A}$ from Eqs. (\ref{inverse1}, \ref{inverse2}) as a function of the number of the bottom-up hierarchical levels for $r=3$ (upper part) and $r=15$ (lower part). In upper green (dark) parts any initial $p_0$ yields at the top of the hierarchy a deterministic victory for A. In lower yellow (light) parts any initial $p_0$ yields at the top of the hierarchy a deterministic failure of A with a victory for B. In the central colored areas  the result at the top is probabilistic. A drastic shrinkage of the probabilistic area is seen when going from $r=3$ (upper part) to $r=15$ (lower part).}
\label{inverses}
\end{figure}

Accordingly, using a bottom-up democratic voting hierarchy instead of a total group voting produces a symmetric shift $\pm \frac{1}{2}\lambda_r^{-n}$ around the $50\%$ democratic threshold. This shift creates two different voting regimes. For $p_{r,n,B}<p_0<p_{r,n,A}$ the voting outcome is probabilistic  and outside this area for $p_0\leq p_{r,n,B}$ and $p_0\geq p_{r,n,A}$ the deterministic outcome of the voting is recovered as seen in in Figure (\ref{inverses}).

The probabilistic area shrinks quite quickly with increasing either the number of levels $n$ or and the size $r$. The extend of the probabilistic surface, which is a function of $n$ and $r$, is measured by 
\begin{eqnarray}
\Delta_{r,n} &\equiv & p_{r,n,A}-p_{r,n,B} \ ,
\nonumber\\
&=& \lambda_r^{-n}  \ .
\label{inverse-ampli}
\end{eqnarray}
Given $r$ its maximum value $\Delta_{r,1}= 1/ \lambda_r$ is reached at $n=1$  since $\lambda_r >1$  with $\lambda_r^{-n}  \rightarrow 0$  for $n\gg 1$  since always $\lambda_r >1$. In addition $\lambda_r \rightarrow \sqrt {\frac{2r}{\pi}}$ for $r\gg 1 \Longrightarrow$
$\Delta_{r,n}\rightarrow  (\frac{2r}{\pi})^{-n/2} \rightarrow 0$  for $r\gg 1$ with any value $n$. 

At this stage Eqs. (\ref{inverse1}) and (\ref{inverse2}) are approximate formulas which result from a Taylor expansion in the vicinity of the unstable fixed point $p_c=\frac{1}{2}$. We thus have to check their accuracy as done above with $n_{c,r}$ given by Eq. (\ref{nc-+}). In this case we had to eventually modify $n_{c,r}$ to $\bar{n}_{c,r}$ from Eq. (\ref{nc-bis}) to get a more accurate formula for predictions. Here the similar correction leads to add a $+2$ correction to the exponent $-n$ as
\begin{equation}
\bar{p}_{r,n,(A,B)} = \frac{1}{2} (1\pm \lambda_r^{-n+2})
\label{lambda-bis}
\end{equation}

The existence of a probabilistic domain implies a coexistence region where the results of an election may yield some unexpected and sudden shift in the elected leadership at the top of the hierarchy. Within that region, no political orientation is sure of winning, which creates an interesting political situation.  For the range $\Delta_{r,n}$ around $50\%$ the outcome is probabilistic with a slight bias in favor of the orientation which is larger, i.e., the current majority. To win with certainty requires to have a little more than just being above $50\%$.

This coexistence region shrinks as a power law $\lambda_r^{-n}$ whose exponent is the number n of hierarchical levels. Therefore, having a small number of hierarchical levels puts higher the threshold to a certain reversal of power since the current minority will need to reach more than fifty percent in order to win for certain the presidency. Simultaneously it lowers the threshold from which  the minority, still being minority, starts to have a non zero chance to win against the current majority as illustrated in Figure (\ref{inverses}). For instance, given $n=5$ we get $0.500\pm 0.148,\pm 0.076,\pm 0.048,\pm 0.016$ for respectively $r=3,5,7,15$.

\section{Rare antidemocratic bottom configurations}

Given a $n$-$brick_r$, associated bottom configurations yielding an A president obtained via a random selections of $r^n$ agents can be discriminated between two families. First one includes configurations which have a A bottom majority while second one regroups configurations for which the number of A is minority, i.e., less than $\frac{r+1}{2}$. Those last configurations thwart the current bottom majority leading to a president who belongs to the bottom minority.

\subsection{Case $r=3$ and $n=2$}
\begin{figure}
\centering
\includegraphics[width=1\textwidth]{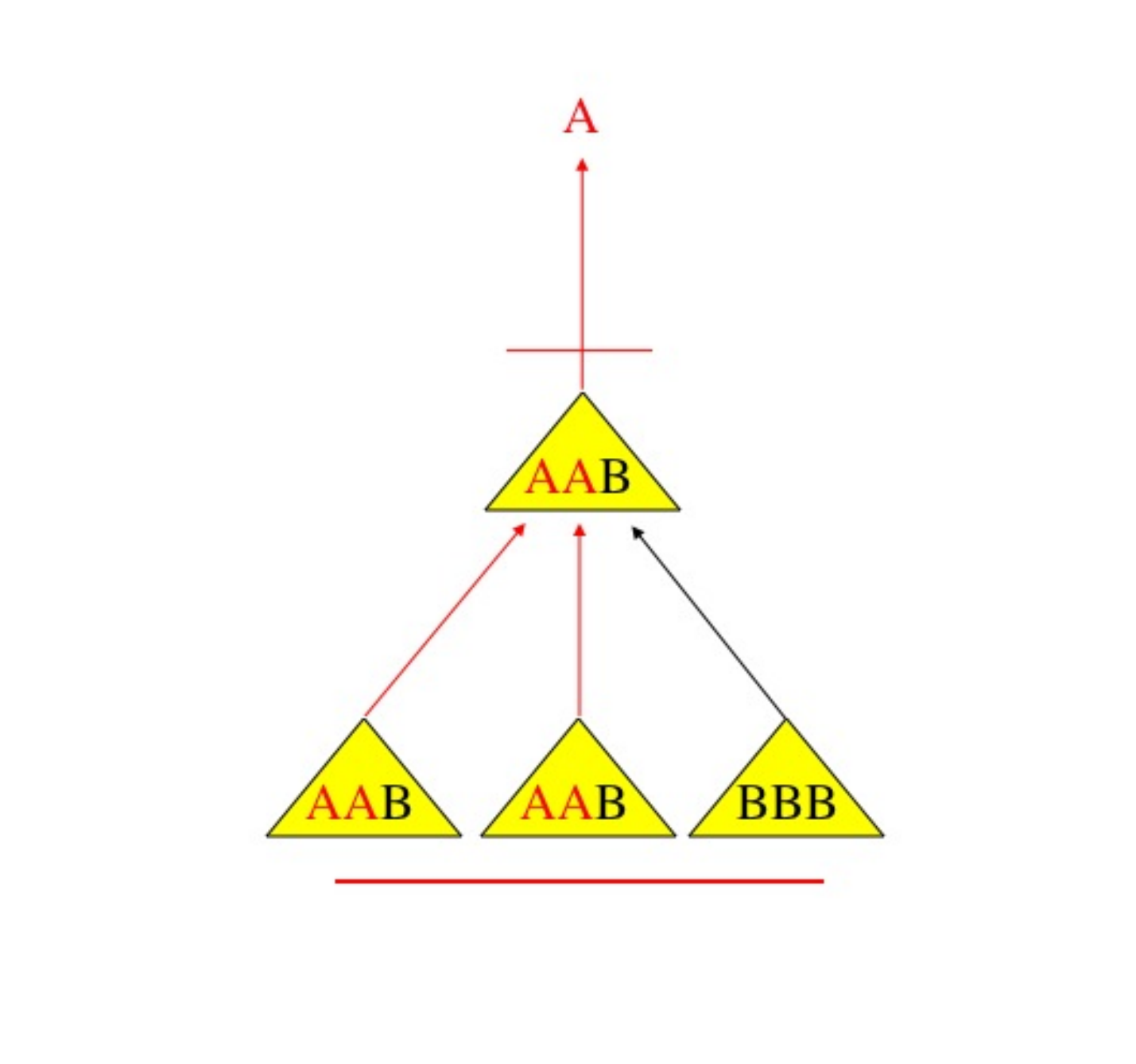}
\caption{ An antidemocratic bottom configuration with a minority number of A-agents $\{4A, 5 B\}$, which yields an A president.  }
\label{anti1}
\end{figure}

An illustration of  an such antidemocratic bottom configuration is shown  in Figure (\ref{anti1}) for $r=3$ and $n=2$. The 4 A-agents present at the bottom against a majority of 5 B-agents obey a $(2, 2, 0)$ distribution of A among the 3 bottom voting groups of size 3. By permutation 27 different  antidemocratic rearrangements can be obtained leading to a probability 
\begin{equation}
P_{3,2}^{4/5}(p_0)=27 p_0^4(1-p_0)^5 ,
\label{anti-a}
\end{equation}
to get such an antidemocratic situation.

This ratio of 4 against 5 is the unique one which can makes the bottom minority to win the presidency. Four agents is the minimum number of agents to achieve this process. Three are not enough and five are already the majority. These figures are a function of both $r$ and $n$.

\begin{figure}
\centering
\includegraphics[width=.8\textwidth]{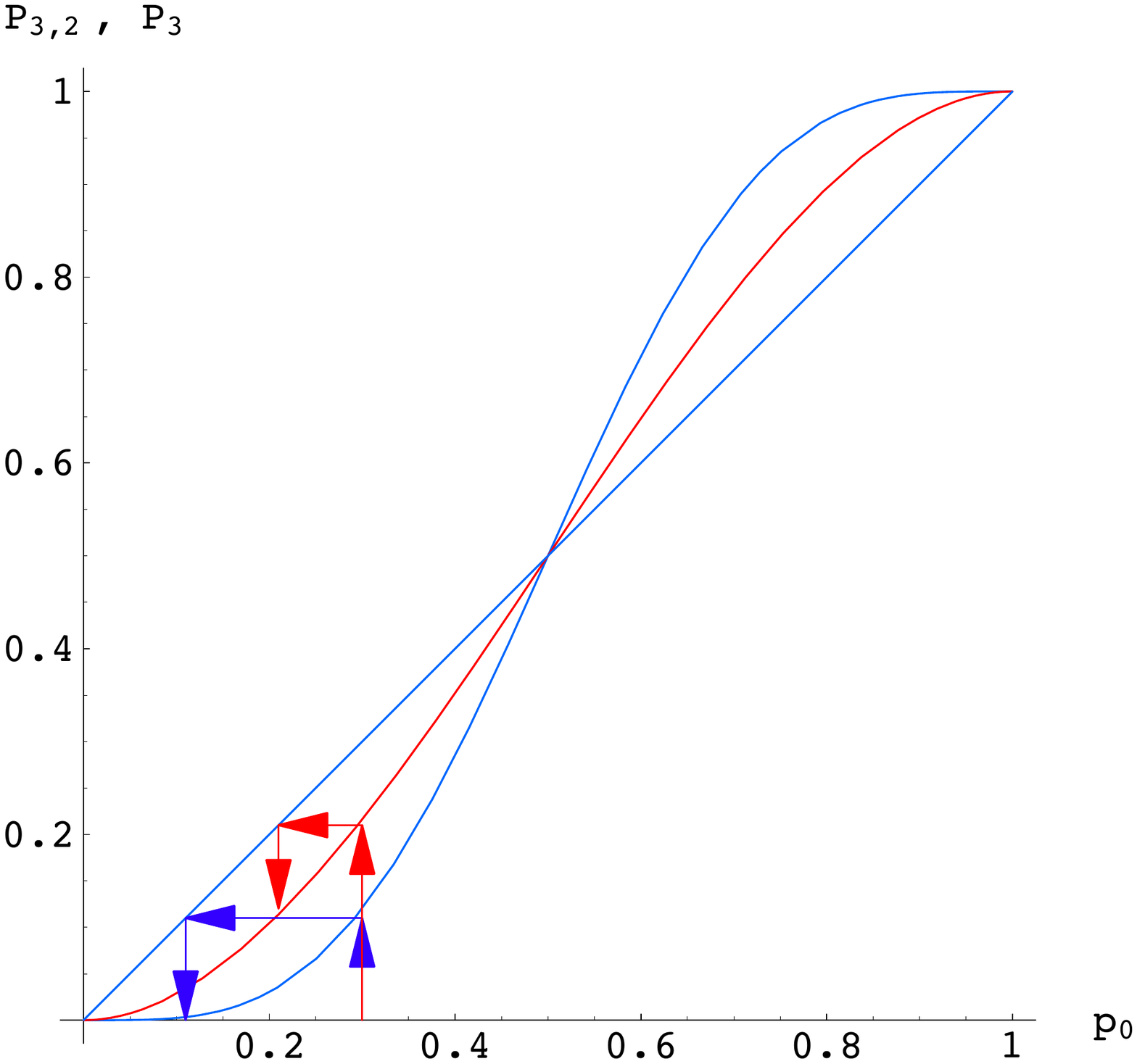}
\caption{ Both probabilities $P_3(p_0)$ and $P_{3,2}(p_0)$ as a function of $p_0$ with the first one being contained within the second one. Two iterations on the inside curve are identical to one iteration on the external curve.}
\label{anti3}
\end{figure}

Going further along this case we can enumerate all configurations which lead to a A victory. In addition to above antidemocratic configurations, we have A democratic configurations with $\{2, 2, 1\}$, $\{3, 2, 0\}$ for a ratio 5 to 4, $\{3, 3, 0\}$, $\{2, 2, 2\}$, $\{3, 2, 1\}$ for a ratio 6 to 3, $\{3, 3, 1\}$, $\{3, 2, 2\}$ for a ratio 7 to 2, $\{3, 3, 2\}$ or a ratio 8 to 1, $\{3, 3, 3\}$ or a ratio 9 to 0 and all associated permutations. Adding all of them gives
\begin{eqnarray}
P_{3,2}(p_0) & = & 27 p_0^4(1-p_0)^5+ 99 p_0^5(1 - p_0)^4 + 84 p_0^6(1 - p_0)^3 \nonumber \\
 & + & 36 p_0^7(1 - p_0)^2 + 9 p_0^8(1 - p_0)^1 + p_0^9 ,
\label{anti-a-majority}
\end{eqnarray}
which is found as expected to be identical to $P_3\{P_3(p_0)\}$. Figure (\ref{anti3}) exhibits both probabilities $P_3(p_0)$ and $P_{3,2}(p_0)$ as a function of $p_0$ with the first one being contained within the second one. 

These findings shed a light on the effect of fragmentation of a single group into smaller subgroups with the simultaneous building of a bottom-up hierarchy.  Looking at all  wining configurations from a single voting group of size $3\times 3=9$ yields
\begin{equation}
P_9(p_0)=p_0^9 + 9 p_0^8 (1-p_0) + 36 p_0^7 (1-p_0)^2 + 84 p_0^6 (1-p_0)^3 + 126 p_0^5 (1-p_0)^4 .
\label{anti9}
\end{equation}
Comparing Eq. (\ref{anti9}) to Eq. (\ref{anti-a-majority}) shows that all coefficients associated to configurations $\{9A, 0B\}, \{8A,1B\}, \{7A,2B\}, \{6A, 3B\}$ are identical. Differences arise from both the coefficient of configuration $\{5A, 4B\}$ and the one of $\{4A, 5B\}$. First one is 126 in  Eq. (\ref{anti9}) against 99 in  Eq. (\ref{anti-a-majority}). Second one in  Eq. (\ref{anti-a-majority}) accounts for a contribution from the minority configuration with a coefficient 27.

Last term $27 p_0^4 (1-p_0)^5$ happens to be exactly the one associated to the antidemocratic configurations found above. By symmetry between the two orientations, similar configurations exist as well in favor of B with the possibility of a winning B minority of 4 agents. This is why, on the one hand the coefficient 126 associated to the configuration $\{4A,5B\}$ in  Eq. (\ref{anti9}) is reduced to 126-27= 99 in Eq. (\ref{anti-a-majority}), and on the other hand there exists a symmetrical contribution in favor of A in Eq. (\ref{anti-a-majority}) with the coefficient 27 at the expense of B  from the configurations $\{4A,5B\}$. These numbers and values of these ratios are a function of both $r$ and $n$.

The fragmentation of the group of 9 agents into 3 separate subgroups of 3 agents each has created the possibility of a few bottom minority winner configurations. It hints quite naturally to envision  the possibility of  ``strategic nesting", which is a crucial ingredient to evaluate the fragility of democratic geometries against possible lobbying.

\subsection{Case $r=3$ and $n=3$}

\begin{figure}
\centering
\includegraphics[width=.8\textwidth]{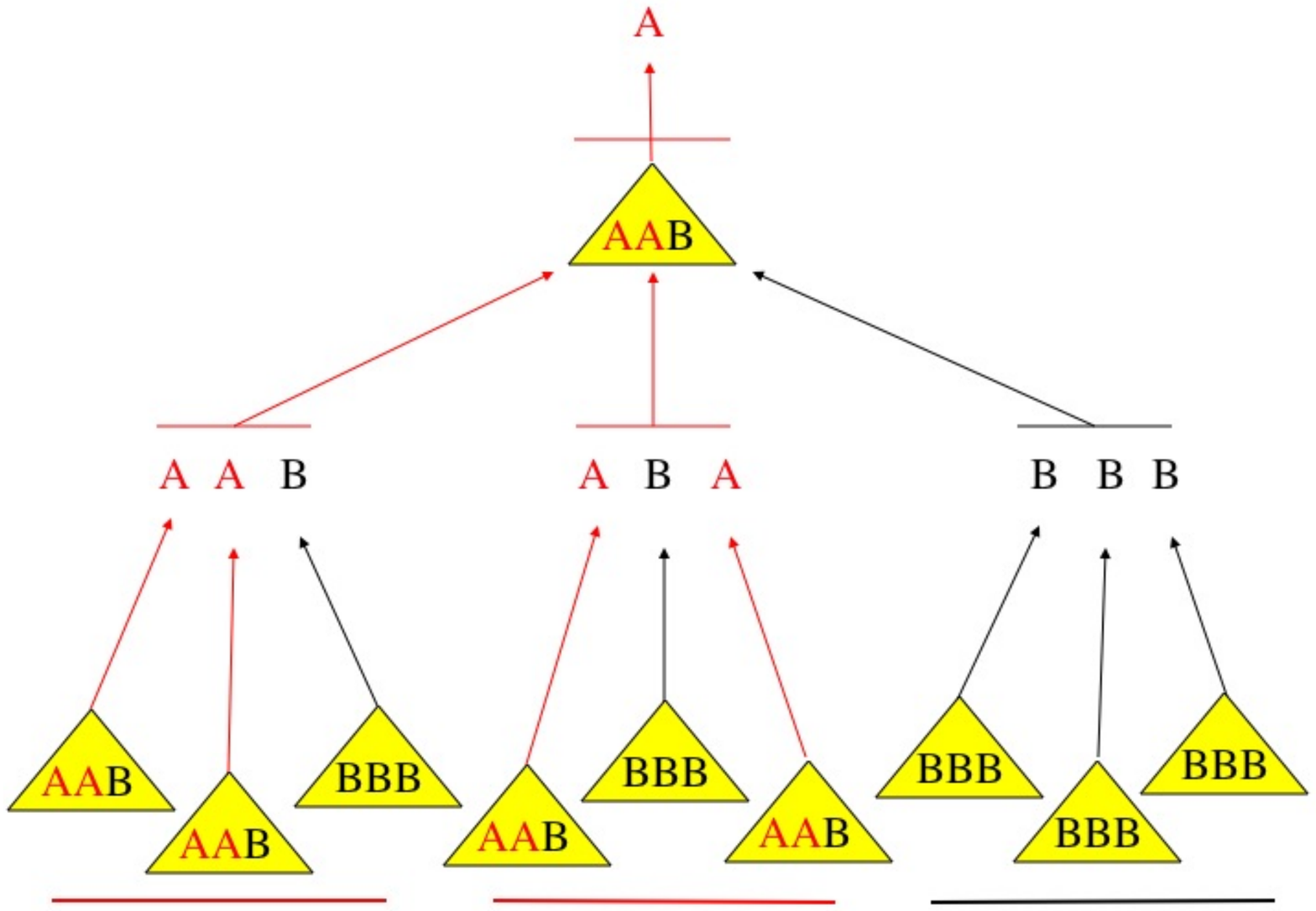}
\caption{ The optimized distribution of 8 A-agents against 19 B-agents within the voting bottom groups of a $r=3, n=3$ hierarchy, which in turn ensures A to win the presidency.}
\label{anti4}
\end{figure}

A $r=3, n=3$ hierarchy is obtained adding one hierarchical level to the $r=3, n=2$ hierarchy treated above. The associated bottom geometry includes now 27 agents distributed among 9 different voting groups of size 3.  Another antidemocratic distribution of  A agents is shown  in Figure (\ref{anti4}).  A minority of 8  A-agents, who are strategically  nested, are sufficient to  hold up the presidency against the majority of 19 B-agents. It is worth to stress that such an antidemocratic bottom configuration can in principle also occur by chance under a random selection of the voting agents. However, as for the precedent case,  the occurrence of such a configuration is a rare event as    it appears from calculating its probability of occurrence.

\begin{figure}
\centering
\includegraphics[width=.8\textwidth]{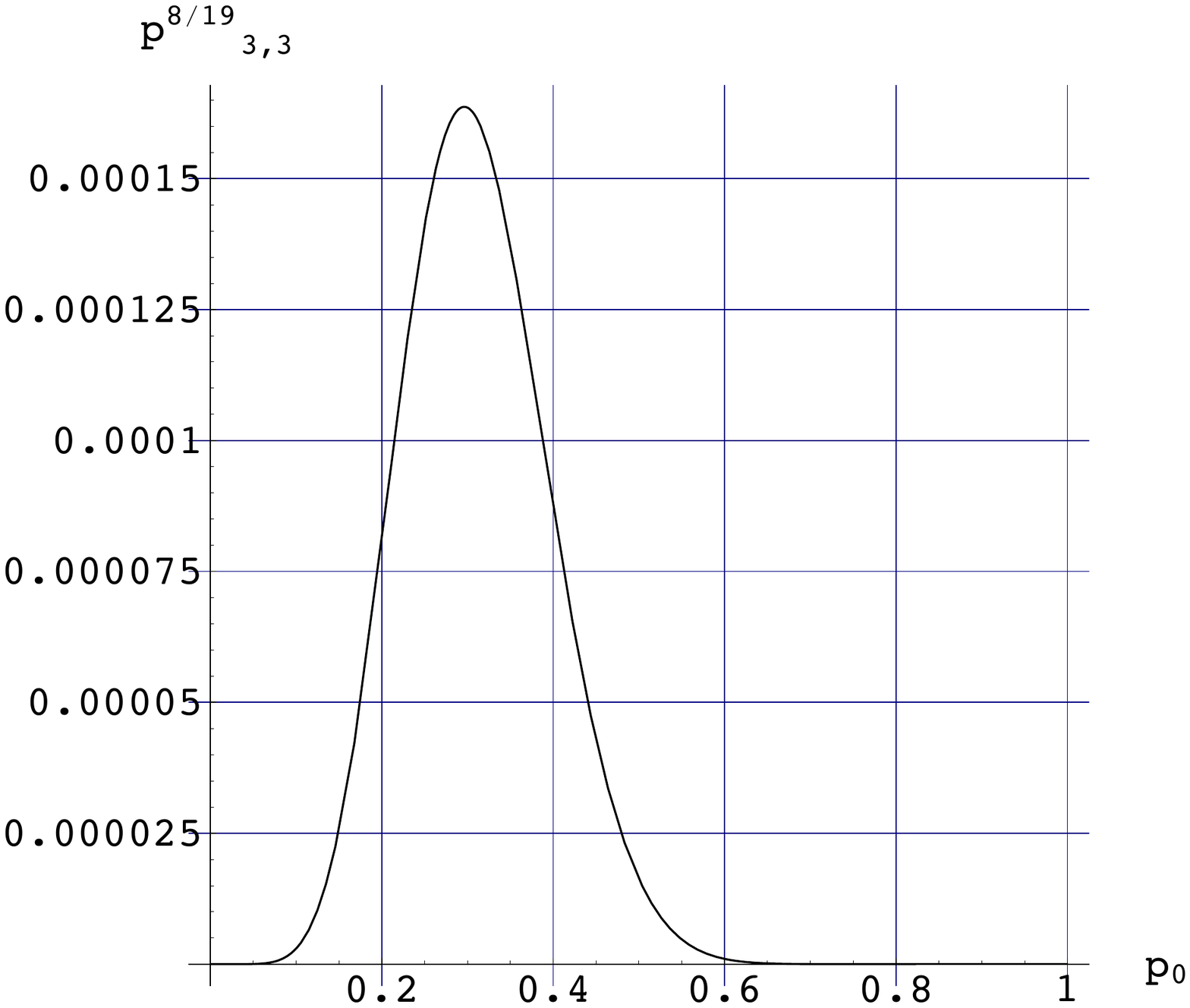}
\caption{ The probability $P_{3,3}^{8/19}(p_0)$ of bottom antidemocratic rearrangements for a $(r=3, n=3)$ hierarchy with the 8-A agents against 19 B-agents (Eq. (\ref{p8/19})). The maximum value is 0.00016 at $p_0=\frac{8}{27}$.}
\label{anti5}
\end{figure}

First contribution comes from the probability to have 8  A-agents and 19 B-agents, it is equal to $p_0^8(1-p_0)^{19}$. Second contribution arises from the number of distinct rearrangements of this minority winning configuration, which scores to $(3 \times 3 \times  3)^2 \times  3=2187$. The final result is
\begin{equation}
P_{3,3}^{8/19} = 2187 p_0^8(1-p_0)^{19} ,
\label{p8/19}
\end{equation}
whose variation as function of $p_0$ is shown in Figure (\ref{anti5}). Its peak reaches the maximum value 0.00016 at $p_0=\frac{8}{27}$ demonstrating how improbable such antidemocratic  bottom configurations are.

Contrary to the precedent case $r=3,n=2$, which has only one antidemocratic distribution (4 against 5) the 8 versus 19 antidemocratic bottom distribution is not unique. It corresponds to the minimum minority size to reach the presidency for a 27 bottom hierarchy. However, there exists many more antidemocratic bottom distributions covering the range from 8/19 up to 13/14, which can yield a democratic victory as the benefit of the bottom minority provided the corresponding agent distributions are antidemocratic, i.e., lead to an A president as in Figure (\ref{anti4}).

\section{One group voting versus small group hierarchies}

To implement the building of our democratic bottom-up hierarchies we have started from one group voting using $r$-$brick_1$ first components. From these elementary bricks one could build higher lego-like bricks to obtain  $r$-$brick_n$. While the voting outcome is probabilistic for these bricks, we saw that some critical number of levels $n_{c,r}$ exists, for which the  $r$-$brick_{n_{c,r}}$ outcome is deterministic (within some number of digits rounding).

In addition to restoring a deterministic voting outcome the hierarchical geometry has the benefit to incorporate representatives from the minority at various levels of the hierarchy, Power is distributed locally involving minority agents although more likely at lower levels than toward the top level. It thus creates a more resilient political structure in which the minority is a stakeholder.

However, comparing Eq. (\ref{anti-a-majority}) and Eq. (\ref{anti9}) we also found that fragmentation into small groups allows the appearance of antidemocratic configurations, yet they are very rare. It is thus of interest to compare the voting outcome from a $r$-$brick_n$ to the one obtained from a $r^n$-$brick_1$, which is a one voting group including all the agents distributed into the $r^{n-1}$ bottom $r$ size groups of the $r$-$brick_n$. The two voting probabilities to compare are thus
\begin{equation}
p_n= P_r(P_r(P_r(...P_r(p_0) ) ) )  ,
\label{pr-odd-n} 
\end{equation}
where $P_r$ is iterated $n$ time given by Eq. (\ref{pr-odd}) and
\begin{equation}
 pp_n\equiv P_R(p_0) \equiv  \sum_{m= \frac{R+1}{2} }^{R}  {R \choose m} p_0^m  (1-p_0)^{R-m} ,
\label{pr-odd-bottom} 
\end{equation}
where $R\equiv r^n$.

\begin{figure}
\centering
\includegraphics[width=1\textwidth]{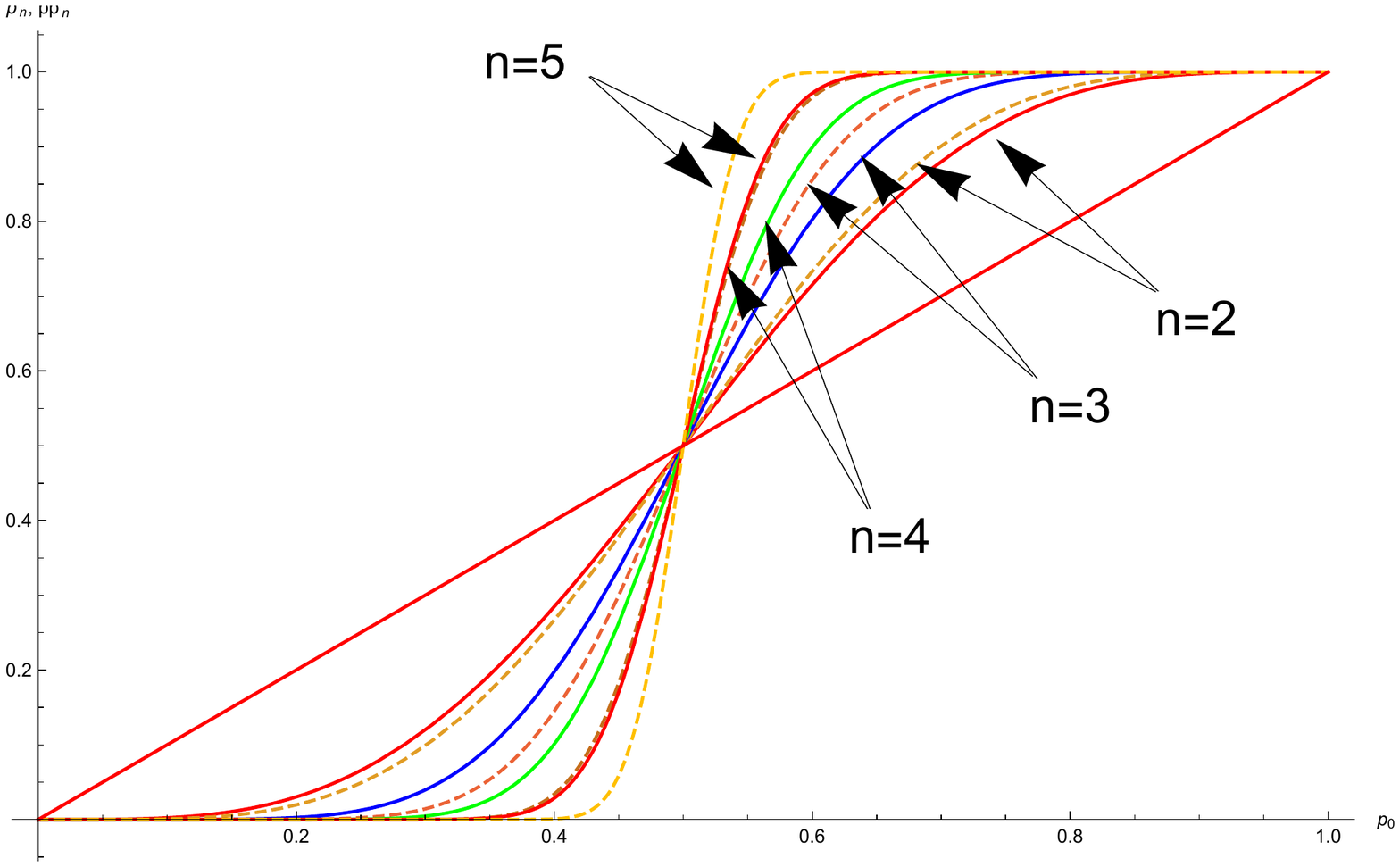}
\caption{Voting probabilities $p_n$ (solid line) and $pp_n$ (dot line) form Eq. (\ref{pr-odd-n}) and Eq. (\ref{pr-odd-bottom}) for the case $r=3$ with $n=2,3,4,5$. Increasing $n$ is seen to enlarge a bit the democratic outperformance of the one group voting. An overlap $p_{n+1} \approx pp_n$ is exhibited for $n=4,5$.}
\label{pn-ppn}
\end{figure}

Above two equations reduces to Eq. (\ref{anti-a-majority}) and Eq. (\ref{anti9}) for $r=3$ and $n=2$ as expected. Figure (\ref{pn-ppn}) shows the case $r=3$ with $n=2,3,4,5$. It is seen that increasing $n$ enlarge a bit the democratic outperformance of the one group voting. This advantage came from the fact that the iteration process allows for a few antidemocratic bottom configurations, which are not present in the bottom one group voting scheme. Therefore increasing $n$ allows for more possible antidemocratic bottom configurations as illustrated in comparing above $r=3$ and $n=2$ versus $r=3$ and $n=3$ cases.  However, a direct voting from the bottom taking as one voting group decreases the number of configurations yielding a democratic outcome yet allowing bottom majorities although they belong to the currently minority in the full population.

\begin{figure}
\centering
\includegraphics[width=1\textwidth]{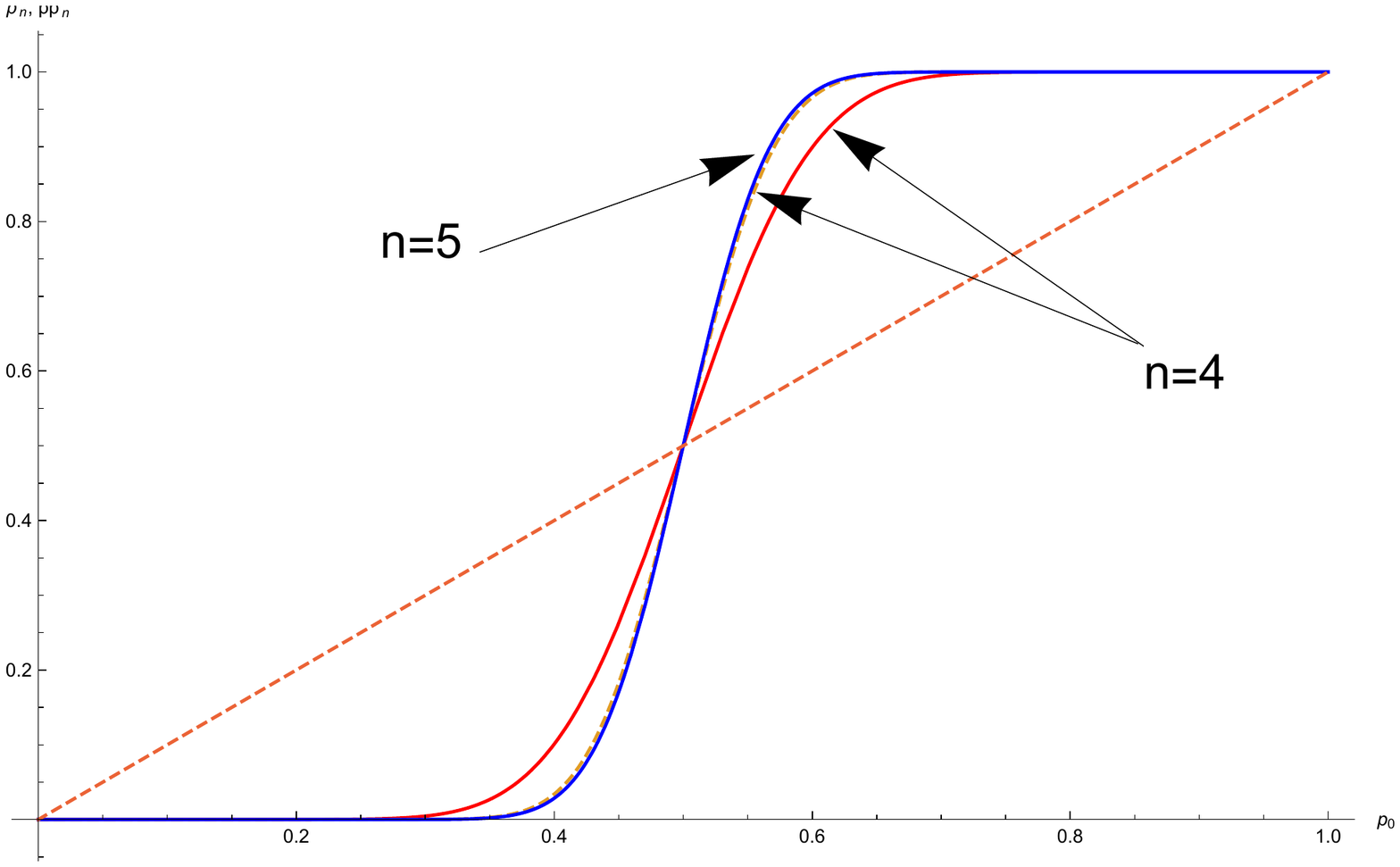}
\caption{ Cases $p_4$ and $p_5$ (solid line) with $pp_4$ (dot line). The overlap $p_{5} \approx pp_4$ is clealy seen.}
\label{n-4-5}
\end{figure}

Interesting to notice from Figure (\ref{n-4-5})  that $p_5 \approx pp_4$ for $r=3$. Such adequacy holds also for larger value of $n$ and other values of $r$. These results hints at a possible quasi-identity $p_{n+1} \approx pp_n$ for $n\geq 4$. However the eventual demonstration is out of scope of the present paper.

 In addition it is of importance to underline that $p_n$ from Eq. (\ref{pr-odd-n}) includes the antidemocratic bottom configurations at the benefit of A but it also excludes by symmetry the antidemocratic bottom configurations with A majority yet at the benefit of B. This fact can be illustrated by evaluating the difference $p_n-pp_n$ for above case $r=3$ and $n=2$. From Eq. (\ref{anti-a-majority}) and Eq. (\ref{anti9}) we get
\begin{equation}
 p_2-pp_2=27 p_0^4(1-p_0)^5 -27 p_0^5(1-p_0)^4,
\label{anti-a-b} 
\end{equation}
where the first term corresponds to Eq. (\ref{anti-a}) and the second term accounts for the symmetric B antidemocratic bottom configurations.

\section{From rare antidemocratic configurations to the killing efficiency of geometric lobbying}

The discovery and identification of rare antidemocratic configurations opens the path to design killing strategies to take over democratic institutions using a combination of democratic rules and instrumental positioning of agents at regular slots. These spots are turned to killing slots only once they get correlated with targeted occupation by a lobbying group. Once strategically located within appropriate bottom voting groups, the ongoing repeated bottom-up local majority rule voting yields the presidency to the bottom lobbying group.

A global monitoring of a given nesting group is thus found to transform above antidemocratic extremely rare events into a certain deterministic outcome. Instead of requiring a support of more than $50\%$ to win the presidency a few agents are now sufficient to hold up the president without any breaking law behavior. Once critically located at the bottom, required minimum numbers of upper level representatives are automatically generated by repeated democratic elections till the presidency. 

\begin{figure}
\centering
\includegraphics[width=1\textwidth]{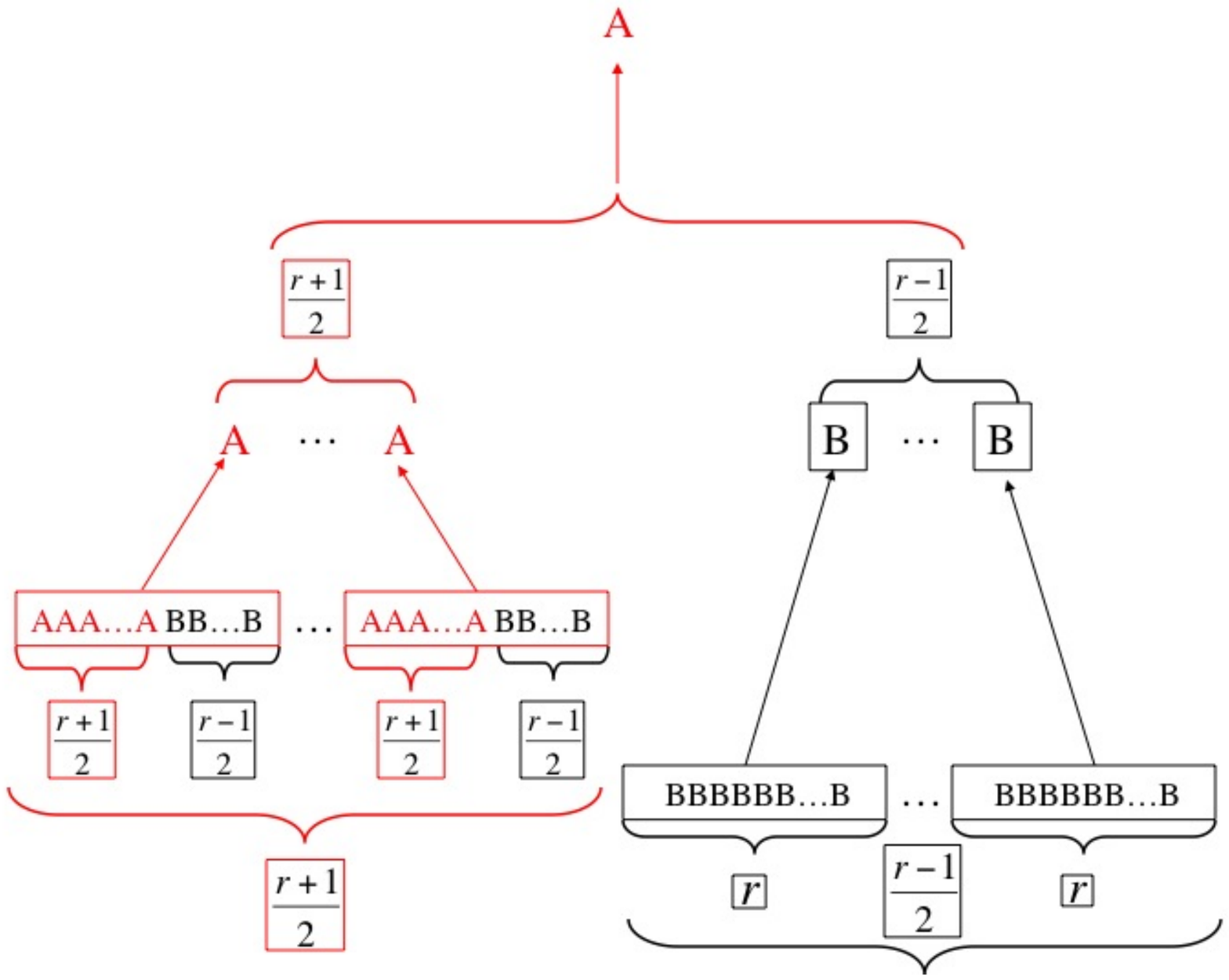}
\caption{ An  antidemocratic bottom configuration for a $n=2$ hierarchy with voting groups of size $r$ featuring the case of the minimum number of A-agents to win the presidency.}
\label{nasty1}
\end{figure}

To proceed, we extend above case $r=3$ and $n=2$ to the general case $r$ and $n$. To get an antidemocratic configuration for $n=1$ requires to have no more but at least $\frac{r+1}{2}$ agents of the lobbying orientation in the unique bottom group of size $r$. It is a $r$-$brick_1$. To move to $n=2$ requires to add one level, i.e., $r$  $r$-$brick_1$. However, only  $\frac{r+1}{2}$ $r$-$brick_1$ are requested to have $\frac{r+1}{2}$ agents to reach the presidency. It scores to a total number $k_{r,2}=\frac{r+1}{2} \times \frac{r+1}{2}= (\frac{r+1}{2})^2$ of nested agents. The corresponding scheme is shown in Figure (\ref{nasty1}). 

Therefore for a  $r$-$brick_n$ a number $(\frac{r+1}{2})^{n-1}$ of $r$-$brick_1$ needs to have at least $\frac{r+1}{2}$ lobbying agents each.The total number of bottom nested agents is
\begin{equation}
 k_{r,n}=(\frac{r+1}{2})^n ,
\label{K} 
\end{equation}
for a total number $r^n$ bottom agents including the nested ones. The killing proportion of nested agents is thus 
\begin{equation}
\bar{k}_{r,n}= \frac{k_{r,n}}{r^n}=\frac{1}{2^n} \left (1+\frac{1}{r}\right )^n ,
\label{k} 
\end{equation}
which can be approximated as
\begin{equation}
\bar{k}_{r,n}\approx \frac{1}{2^n} \left (1+\frac{n}{r}\right ).
\label{kk}
\end{equation}

\begin{figure}
\centering
\includegraphics[width=.8\textwidth]{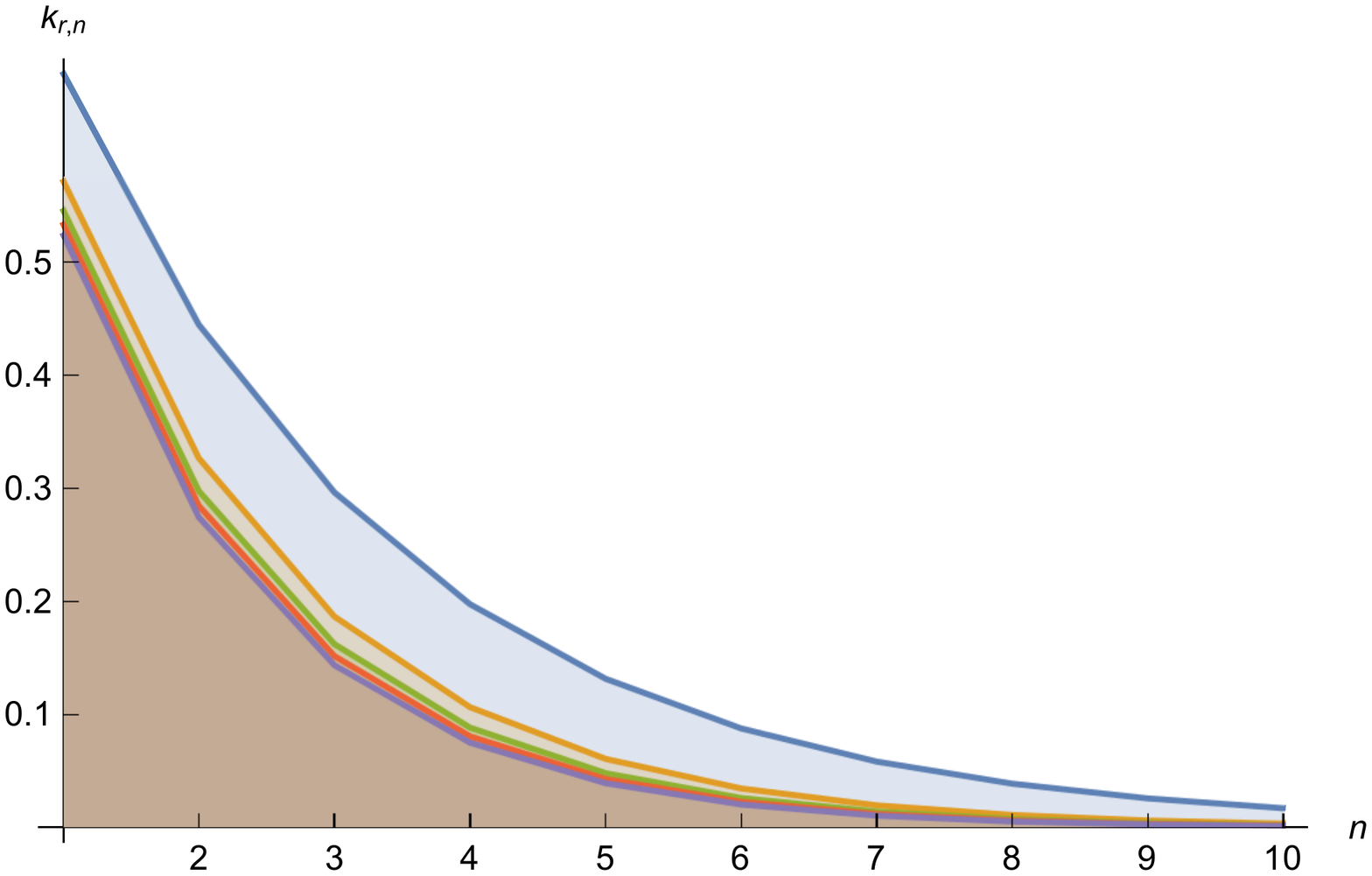}
\caption{The evolution of $k_{r,n}$ from Eq. (\ref{K}) for 5 values of the voting size groups  $r=3,7,11,15,21$ as a function of the number of levels $n$.  }
\label{nasty2}
\end{figure}

\begin{figure}
\centering
\includegraphics[width=.8\textwidth]{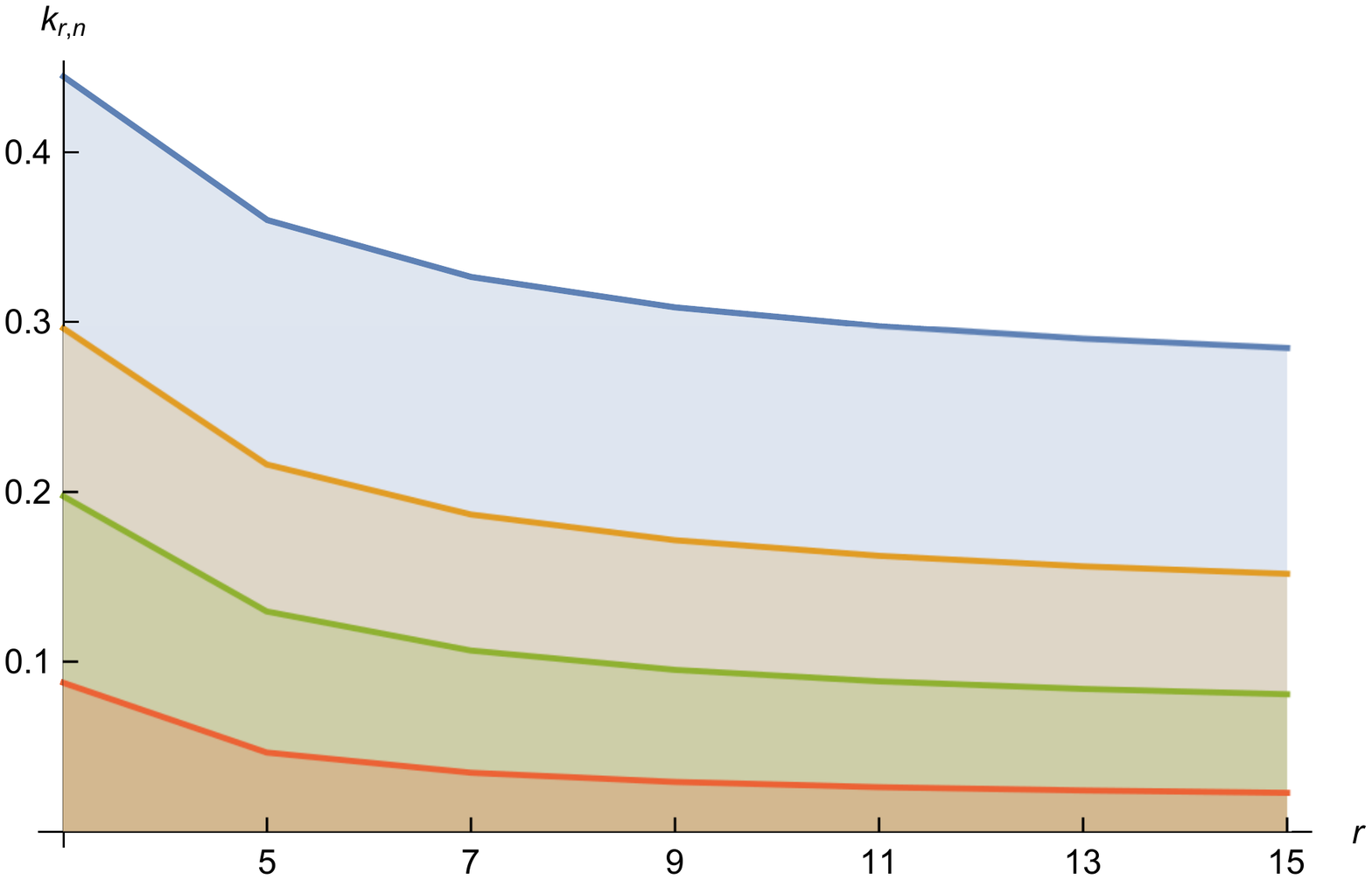}
\caption{ The evolution of $k_{r,n}$ from Eq. (\ref{K}) for 4 values of the number of levels with $n=2,3,4,6$ as a function of the voting size groups $r$.}
\label{nasty3}
\end{figure}

Figure (\ref{nasty2}) shows $\bar{k}_{r,n}$ from Eq. (\ref{k}) as a function of the number of levels $n$ given a series of fixed values of the voting size group $r=3,7,11,15,21$. The variation of $k_{r,n}$ as a function of the group size $r$ for a fixed number of levels with at $n=2,3,4,6$ is exhibited in Figure (\ref{nasty3}). Both Figures illustrate the  quick and drastic lowering from the democratic threshold $\frac{r^n+1}{2}$ for the requested proportion of agents to win the presidency using a geometric killing nesting. 

\section{The impossible identification of an existing killing geometric lobbying}

At this stage it is worth to underline that a killing bottom configuration can be modified by performing all permutations within the nested groups and between nested groups and non-nested groups yet achieving the same performance of electing a minority president. The number of permutations associated to a killing bottom configuration increases quite rapidly with $r$ and $n$. This fact makes impossible the identification of an actual implemented targeted lobbying. 

The total number of existing bottom killing configurations can be calculated exactly for a given bottom-up democratic hierarchy from Eq. (\ref{pr-odd-n}). Indeed it is a polynomial expansion in $p_0$ and $(1-p_0)$. The highest degree of $p_0$ is $r^n$ with $r^n$ A and zero B. The lowest degree gives the minimum number of nested agents, i.e., $k_{r,n}$ (Eq. (\ref{K})). We thus have

\begin{equation}
 p_n={p_0}^{r^n}+r^n {p_0}^{r^{n-1}}(1-p_0) +\ldots+g_{r,n} {p_0}^{k_{r,n}}(1-p_0) ^{r^n-k_{r,n}},
\label{k-plus} 
\end{equation}
where $g_{r,n}$ is the minimum number of agents required to seize democratically the presidency. its value can be evaluated exactly as
\begin{equation}
 g_{r,n}=\left(\frac{r!}{\frac{r+1}{2}!\frac{r-1}{2}!}\right)^{\frac{(r+1)^n-2^n}{2^{n-1}(r-1)}}
\label{g} 
\end{equation}
where the exponent is the sum of the geometric series $1+\frac{r+1}{2}+\left(\frac{r+1}{2}\right)^2+\ldots+\left(\frac{r+1}{2}\right)^{n-1}$. It happens that $g_{r,n}$ reaches astronomical values quite immediately. For instance it yields, $27, 2187,14348907$ for respectively $n=2,3,4$ at $r=3$ and $10^4, 10^{13}, 10^{31}$ for respectively $n=2,3,4$ at $r=5$. Accordingly for any hierarchy the number of correlated slots which allow the president seizing is ``infinite" in practical terms.

In addition to the countless number of killing configurations with the minimum number of nested agents, quite more killing geometries are available. Indeed all cases where the presidency is won by a bottom minority suitably distributed yields the identical democratic thwarting. All these cases are directly identified by calculating the difference $\delta_{r,n}\equiv p_n-pp_n$ from Eq. (\ref{pr-odd-n}) and Eq. (\ref{pr-odd-bottom}) as illustrated above for $r=3$ and $n=2$ with Eq. (\ref{anti-a-b}). For a bigger hierarchy with $n=3$ we get
\begin{eqnarray}
\delta_{3,2}&=&2187 p^{8 }(1-p)^{19}+ 35721 p^{9 }(1-p)^{18 } 266085 p^{10 }(1-p)^{17 }\nonumber \\
&& + 1203903 p^{11 }(1-p)^{16 } + 3677346 p^{12 }(1-p)^{15 } +7902198 p^{13 }(1-p)^{14 }\nonumber \\
&&-2187 p^{19 }(1-p)^{8 }- 35721 p^{18 }(1-p)^{9 }- 266085 p^{17 }(1-p)^{10 } \nonumber \\
&& -1203903 p^{16 }(1-p)^{11 } - 3677346 p^{15 }(1-p)^{12 }- 7902198 p^{14 }(1-p)^{13} , \nonumber \\
\label{delta3-2}
\end{eqnarray}
which exhibits the expected symmetry between both orientations A and B. Negative terms account for antidemocratic configurations which are at the benefit of B. The minimum number of killing agents is found to be 8 against 19 with 2187 configurations as evaluated above with $g_{3,3}=2187$ from Eq. (\ref{g}).

\begin{figure}
\centering
\includegraphics[width=.8\textwidth]{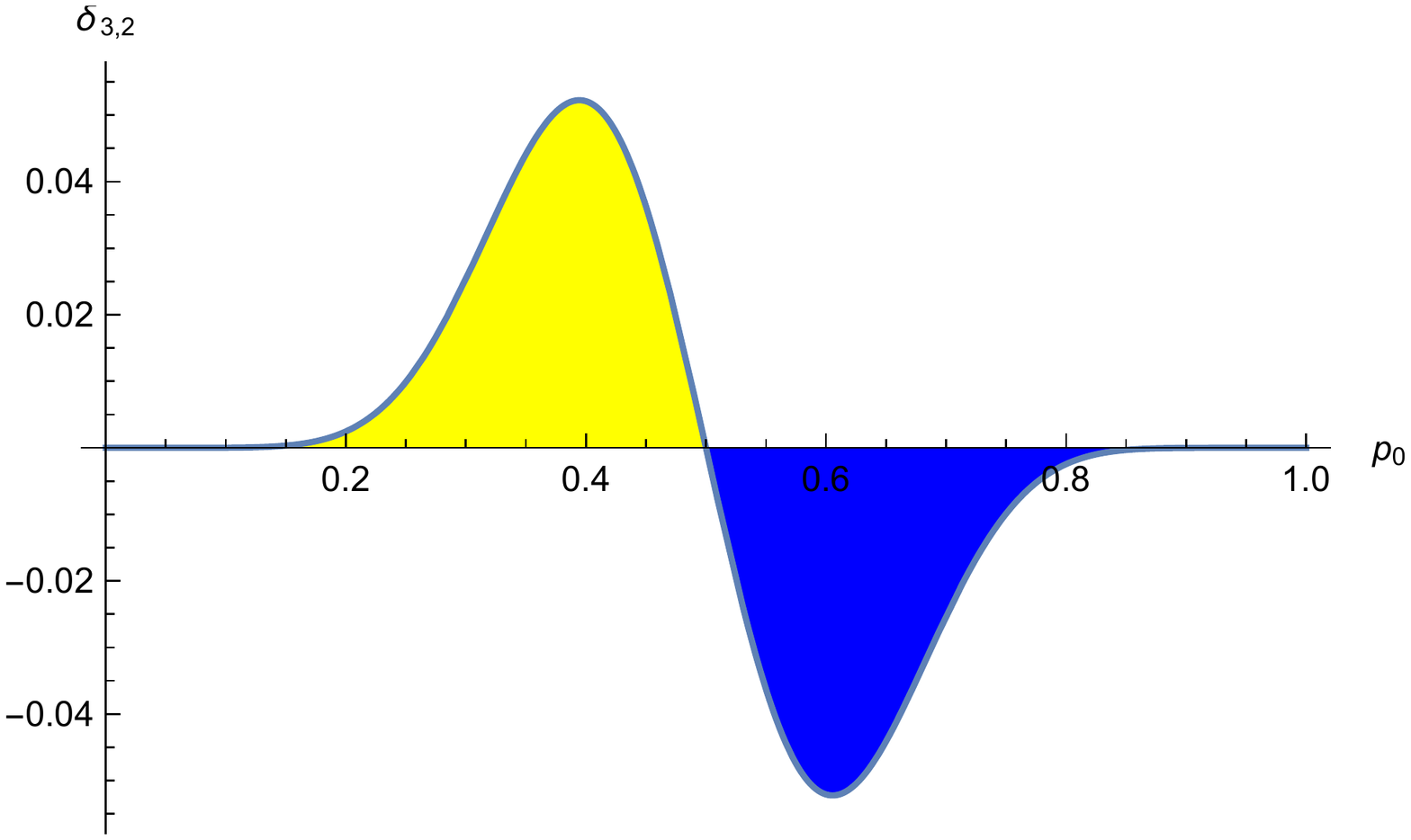}
\caption{Variation of $\delta_{3,2}$ from Eq. (\ref{delta3-2}) as function of $p_0$. Upper part ( $>0$ ) is at the benefit of A while below part ( $<0$ ) is at the expense of A. }
\label{nasty3-2}
\end{figure}

For every killing geometry at the benefit of A exists the same one at the benefit of B. However their respective contributions to the presidency winning probability vary with the proportion $p_0$ of A supporters in the population. Those killing configurations are more beneficial  to the overall minority as is seen in Figure (\ref{nasty3-2}). When $p_0<\frac{1}{2}$ the net contribution between the killing configurations, which yield a A president and the ones which lead to a B president, is positive. It turns negative for $p_0>\frac{1}{2}$.

Last but nor least, the lobbying group does not preclude to have additional supporters randomly distributed among the bottom voting groups, which increases the blurring of the lobbying activity making its detection quasi-impossible. These extra agents do not need to be aware about the ongoing lobbying by their group, contributing further to the screening of 
the on going manipulation.

\section{Conclusion}

We have provided an alternative voting scheme to the universal suffrage election of a president. The scheme is built from elementary bricks of $r$ agents randomly selected from the whole population. Each group elects a local president according to its local majority rule. We shown that the deterministic universal suffrage outcome can be recovered provided enough hierarchical levels are included in the bottom-up democratic hierarchy built from elementary bricks.

In addition to restoring a deterministic voting outcome the hierarchical geometry has the advantage to associate representatives from the minority at various levels of the hierarchy. Power is thus distributed locally involving minority agents although more likely at lower levels than in the top level vicinity. Yet, the top presidency is allocated with certainty to the actual majority in the whole population. The hierarchical structure is thus resilient towards temptation from the minority to call into question the presidency legitimacy.

The scheme produces also an area around fifty percent for which the president is elected with an almost equiprobability slightly biased in favor of the actual majority thus preventing destabilizing controversies when the election outcome is vey narrow around fifty percent,\cite{fifty}.

However, we also found that small groups bottom fragmentation produces antidemocratic configurations. Yet, despite being very rare, these antidemocratic configurations open the path to design killing geometries for a lobbying group. It is sufficient to implement some correlated distribution of agents at a series of bottom slots to win the presidency democratically just applying the democratic majority rule voting at all its hierarchy levels. 

We were thus able to reveal the existence of a severe geometric vulnerability of pyramidal structures against lobbying. Moreover, at the present stage, identifying an actual killing distribution is not feasible, which sheds a disturbing light on the devastating effect geometric lobbying can have on democratic hierarchical institutions.

Of course, our model may sound exaggerated and in many aspects it is. But the purpose is not to describe precisely a real situation. On the contrary, the aim is to exhibit the formal possibilities, which do exist in real organizations, to divert completely a democratic dynamics at the expense of the will of people majority. We have enlightened the advantage an infinitesimal small group can gain in playing the structure against its democracy founding principle. Hopefully further studies will determine possibilities to  thwart these ``geometric coups". For future work it will be also interesting to investigate those killing geometries in presence of $k\neq \frac{1}{2}$, which makes even sizes to depart significantly from odd sizes,\cite{mino}.

\section*{Acknowledgment}
I would like to thank Nicola Bellomo for stimulating discussion about the mathematical potential of sociophysics.

\end{document}